\documentclass[prd,aps,twocolumn,showpacs,preprintnumbers,amsmath,amssymb,superscriptaddress,nofootinbib]{revtex4-1}
\usepackage[dvips]{graphicx}
\usepackage{epsf}
\usepackage{amsmath}
\usepackage{amssymb}
\usepackage{amsfonts}
\usepackage{times}
\usepackage[usenames]{color}
\usepackage[normalem]{ulem}
\usepackage[T1]{fontenc}
\usepackage[dvipsnames]{xcolor}

\usepackage{graphicx}
\usepackage{dcolumn}
\usepackage{bm}

\begin{document}

\title{Weighing neutrinos in $f(R)$ gravity}

\author{Jian-hua He}
\email[Email address: ]{jianhua.he@brera.inaf.it}
\affiliation{INAF-Osservatorio Astronomico, di Brera, Via Emilio Bianchi, 46, I-23807, Merate (LC), Italy}


\begin{abstract}
We constrain the neutrino properties in $f(R)$ gravity using the latest observations from cosmic microwave background(CMB) and baryon acoustic oscillation(BAO) measurements. We first constrain separately the total mass of neutrinos $\sum m_\nu$ and the effective number of neutrino species $N_{\rm eff}$. Then we constrain $N_{\rm eff}$ and $\sum m_\nu$ simultaneously. We find $\sum m_\nu<0.462 {\rm eV}$ at a 95\% confidence level for the combination of Planck CMB data, WMAP CMB polarization data, BAO data and high-$l$ data from the Atacama Cosmology Telescope and the South Pole Telescope. We also find
$N_{\rm eff}=3.32^{+0.54}_{-0.51}$ at a 95\% confidence level for the same data set. When constraining $N_{\rm eff}$ and $\sum m_\nu$ simultaneously, we find
$N_{\rm eff}=3.58^{+0.72}_{-0.69}$ and $\sum m_\nu<0.860{\rm eV}$ at a 95\% confidence level, respectively.

\end{abstract}

\maketitle
\section{Introduction}
The determination of the neutrino mass is an important issue in fundamental physics. The Standard Model of particle physics had assumed that all three families of neutrinos: electron neutrinos $\nu_e$, muon neutrinos $\nu_{\mu}$ and tau neutrinos $\nu_{\tau}$ are massless, and that the neutrino cannot change its flavor from one to another. However, the results from solar and atmospheric experiments \cite{experiment} showed that the flavour of neutrinos could oscillate. The mixing and  oscillating of flavors implies nonzero differences between the neutrino masses, which in turn indicates that the neutrinos have absolute mass. If the neutrino does have absolute mass, it will be the lowest-energy particle in the extensions of the Standard Model of particle physics. However, such observations of flavor oscillations can only show that the neutrinos have mass, and cannot exactly pin down the absolute mass scale of neutrinos. Particle physics experiments are able to place lower limits on the effective neutrino mass, which, however, depends on the hierarchy of the neutrino mass spectra\cite{Hitoshi}(also see Ref.\cite{nu_review} for reviews).

On the other hand, cosmological constraints on neutrino properties are highly complementary to
particle physics. Massive neutrinos, if above $1 {\rm eV}$, will become nonrelativistic before recombination\cite{Komatsu}, leaving an impact on the first acoustic peak in the cosmic microwave background(CMB) temperature angular power spectrum due to the early-time integrated Sachs-Wolfe (ISW) effect; neutrinos with mass below $1 {\rm eV}$ will become nonrelativistic after recombination, altering the matter-radiation equality; the massive neutrino will also suppress the matter power spectrum on small scales, since neutrinos
cannot cluster below the free-streaming scales~\cite{waynehu}(see\cite{Julien} for reviews). Combining various cosmological observations can put rather tight constraints on the sum of the neutrino mass. The most recent measurements from the Planck satellite\cite{planck} on the CMB in combination with the baryon acoustic oscillation(BAO)\cite{BAO1,BAO2,BAO3,BAO4}, WMAP polarization(WP) and the high-$l$ data on the CMB from the Atacama Cosmology Telescope(ACT)\cite{ACT} and the South Pole Telescope(SPT)\cite{SPT} give an upper limit for the sum of the neutrino mass as $\sum m_{\nu}<0.23{\rm eV}(95\% {\rm {C.L.}})$ in the spatially flat $\Lambda\rm{CDM}$ model with the effective number of neutrino species as $ N_{\rm eff} = 3.04$. It is even more promising that with the upcoming ESA Euclid mission\cite{Euclid} in the near future, the neutrino mass can be constrained up to an unprecedented accuracy simply by cosmological observations\cite{luca}. The allowed neutrino mass window could be closed by forthcoming cosmological observations.

Nevertheless, it is important to recall that the constraints on neutrino properties are usually found within the context of a $\Lambda$CDM model or within the context of a dark energy model\cite{neutrinolcdm}. Considering different cosmological models, degeneracies may arise among neutrinos and other cosmological parameters.  Cosmological constraints on neutrino properties are highly model dependent. References\cite{luca,Giusarma} have investigated this issue in the framework of a dark energy model with varying total neutrino mass and number of relativistic species. The aim of this paper is, however, to extend such investigations to modified gravity models.  For simplicity, we consider the $f(R)$ gravity~\cite{fr} and particularly focus on a specific family of $f(R)$ models that can exactly reproduce the $\Lambda$CDM background expansion history of the Universe. This family of $f(R)$ models has only one more parameter than the $\Lambda$CDM model, which can be characterized by
\begin{equation}
B_0=\frac{f_{RR}}{F}\frac{dR}{dx}\frac{H}{\frac{dH}{dx}}(a=1)\quad,
\end{equation}
which is approximately the squared Compton wavelengths in units of
the Hubble scale~\cite{Song}. Cosmological constraints on these models without taking into account neutrino mass have already been presented in the literature. On linear scales, the WMAP nine-year data in combination with the matter power spectra of LRG from SDSS DR7 data can only put weak constraints on these models: $B_0<3.86({\rm 95\%C.L.})$\cite{frlinear,songfitting}. Tighter constraints can be obtained from the galaxy-ISW correlation data, which puts the constraint up to $B_0<0.376({\rm 95\%C.L.})$\cite{frlinear,Lucas}. Using the data of cluster abundance, the constraints are dramatically improved up to $B_0<1.1\times 10^{-3}({\rm 95\%C.L.})$ \cite{Lucas,schmidt}. However, the tightest constraints so far come from the astrophysical tests\cite{Jain} which place the upper bound for $B_0$ as $B_0<2.5\times10^{-6}$. On the other hand, the cosmological constraints on $f(R)$ models taking into account neutrino mass have also already been presented in the literature\cite{neutrinofr,gbz}. However, these works are done within the framework of parameterized gravities. We still need to get more accurate results by solving the full linear perturbation equations in the $f(R)$ gravity.

In this paper, we will explore the neutrino properties in $f(R)$ gravity based on our modified version of CAMB code~\cite{CAMB}, which solves the full linear perturbation equations in the $f(R)$ gravity~\cite{frlinear}. We will conduct the Markov chain Monte Carlo(MCMC) analysis on our model based on the COSMOMC package\cite{mcmc} and constrain the cosmological parameters using the latest observational data. Besides examining the total mass of active neutrinos $\sum m_{\nu}$, we will also investigate the effective number of neutrino species $N_{\rm eff}$ since a detection of $N_{\rm eff} > 3.04$ will imply additional relativistic relics or nonstandard neutrino properties\cite{mangano}.

This paper is organized as follows: in section \ref{frgravity}, we will briefly outline the details of the basic equations in $f(R)$ cosmological models. In section \ref{ISW_lensing}, we will discuss about how the $f(R)$ gravity impacts on the neutrino constraints.
In Sec. \ref{Data}, we will list the observational data used in this work. In Sec. \ref{Num}, we will present the details of our numerical results. In Sec. \ref{conclusions}, we will summarize and conclude this work.
\section{$f(R)$ gravity\label{frgravity}}
In $f(R)$ gravity, the Einstein-Hilbert action is given by
\begin{equation}
S=\frac{1}{2\kappa^2}\int d^4x\sqrt{-g}f(R)+\int d^4x\mathcal{L}^{(m)}\quad,
\end{equation}
where $\kappa^2=8\pi G$ and $\mathcal{L}^{(m)}$ is the matter Lagrangian. With variation with respect to $g_{\mu\nu}$, we obtain the modified Einstein equation
\begin{equation}
FR_{\mu\nu}-\frac{1}{2}fg_{\mu\nu}-\nabla_{\mu}\nabla_{\nu}F+g_{\mu\nu}\Box F=\kappa^2T_{\mu\nu}^{(m)}\quad ,\label{FRfield}
\end{equation}
where $F=\frac{\partial f}{\partial R}$. If we consider a homogeneous and isotropic background universe described by the flat Friedmann-Robertson-Walker(FRW) metric
\begin{equation}
ds^2=-dt^2+a^2dx^2\quad,
\end{equation}
the modified Friedmann equation in $f(R)$ gravity is given by\cite{frreview}
\begin{equation}
H^2=\frac{FR-f}{6F}-H\frac{\dot{F}}{F}+\frac{\kappa^2}{3F}\rho\label{field}\quad.
\end{equation}
Taking the derivative of the above equation, we obtain
\begin{equation}
\ddot{F}+2F\dot{H}-H\dot{F}=-\kappa^2(\rho+p)\quad,\label{dfield}
\end{equation}
where the dot denotes the time derivative with respect to the cosmic time $t$, and $\rho $ is the total energy density of the matter which consists of
the cold dark matter, baryon, photon, and neutrinos. $p$ is the total pressure in the Universe.
If we convert the derivatives in Eq.(\ref{dfield}) from the cosmic time $t$ to $x=\ln a$ ,
Eq.(\ref{dfield}) can be written as
\begin{equation}
\frac{d^2}{dx^2}F+(\frac{1}{2}\frac{d\ln E}{dx}-1)\frac{dF}{dx}+(\frac{d\ln E}{dx})F=\frac{\kappa^2}{3E}\frac{d\rho}{dx}\quad ,\label{Ffield}
\end{equation}
where $E\equiv\frac{H^2}{H_0^2}$ and $\frac{d\rho}{dx}=-3(\rho+p)$. For convenience, in the above equation, the energy density $\rho$ is in units of $H_0^2$, and we set $\kappa^2=1$ in our analysis. In order to mimic the $\Lambda$CDM background expansion history, we can parameterize $E(x)$ as \cite{WMAP}
\begin{equation}
E(x)=(\Omega_c^0+\Omega_b^0)e^{-3x}+\Omega_d^0+\Omega_r^0e^{-4x}[1+0.227N_{\rm eff}f(m_{\nu}e^{x}/T_{v0})]\quad,
\end{equation}
which includes the effect of neutrinos. $\Omega_c^0$ and $\Omega_b^0$ represent present-day  cold dark matter and baryon density, respectively. $\Omega_d^0$ is the effective dark energy density which is a constant. $T_{\nu0}=(4/11)^{1/3}T_{\rm cmb}=1.945{\rm K}$ is the present-day neutrino temperature and $\Omega_r^0=2.469\times10^{-5}h^{-2}$ for $T_{\rm cmb}=2.725{\rm K}$. $m_{\nu}$ represents the neutrino mass and we assume that all massive neutrino species have the equal mass. The function $f(y)$ in the above expression is defined by
\begin{equation}
f(y)=\frac{120}{7\pi^4}\int_0^{+\infty}dx\frac{x^2\sqrt{x^2+y^2}}{e^x+1}\quad.
\end{equation}
After fixing the background expansion, Eq.(\ref{Ffield}), governing the behavior of the scale field $F(x)$ in $f(R)$ gravity, can be solved numerically, given the initial condition in the deep-matter-dominated epoch\cite{frlinear}:
\begin{equation}
\begin{split}
F(x)&\sim1+D(e^{3x})^{p_+}\quad,\\
\frac{dF(x)}{dx}&\sim3Dp_+(e^{3x})^{p_+}\quad,\label{initial}
\end{split}
\end{equation}
where the index is defined by $p_+=\frac{5+\sqrt{73}}{12}$. The above initial conditions are still applied here, because the relativistic neutrinos are far less than the total amount of nonrelativistic species(including baryons, cold dark matter and nonrelativistic neutrino)in the Universe at this moment. Equation (\ref{Ffield}) has analytical solutions\cite{frmodel} if we ignore the relativistic species in the Universe. Noting the fact that $p_+>0$, our model only has growing modes in the solutions of Eq.(\ref{Ffield}), which satisfy
\begin{equation}
\lim_{x\rightarrow-\infty}F(x)=1\quad,
\end{equation}
and our model thus can go back to the $\Lambda$CDM model at high redshift.

This family of $f(R)$ models has only one more parameter than the $\Lambda$CDM model, which can be characterized either by $D$ or by the Compton wavelengths $B_0$.
In this work, we will sample $D$ directly in our MCMC analysis and treat $B_0$ as a derived parameter. In order to avoid the instabilities in the high-curvature region\cite{Ignacy}, we need to set $D<0$, which keeps the Compton wavelength $B$ always positive during the past expansion of the Universe $B>0$.

We set the initial conditions for the background in Eq.(\ref{dfield}) roughly at the point $a_i\sim0.03$ around which the value of the scalar field $F(x)$ obtained by solving Eq.(\ref{dfield})rather weakly depends on the exact choice of $a_i$ , given Eq(\ref{initial}) as the initial conditions. For the perturbed spacetime, we solve the full linear perturbation equations in the $f(R)$ gravity based on our modified version of the CAMB code~\cite{frlinear}. In our code, we plug in the $f(R)$ gravity perturbation at $a=0.03$, before which we set the perturbation as $\delta F=0$,$\dot{\delta F}=0$ such that the equations completely go back to the standard equations in the $\Lambda$CDM model.
\section{The Integrated Sachs$-$Wolfe effect and the CMB lensing\label{ISW_lensing}}
\begin{figure}
\includegraphics[width=3in,height=2.8in]{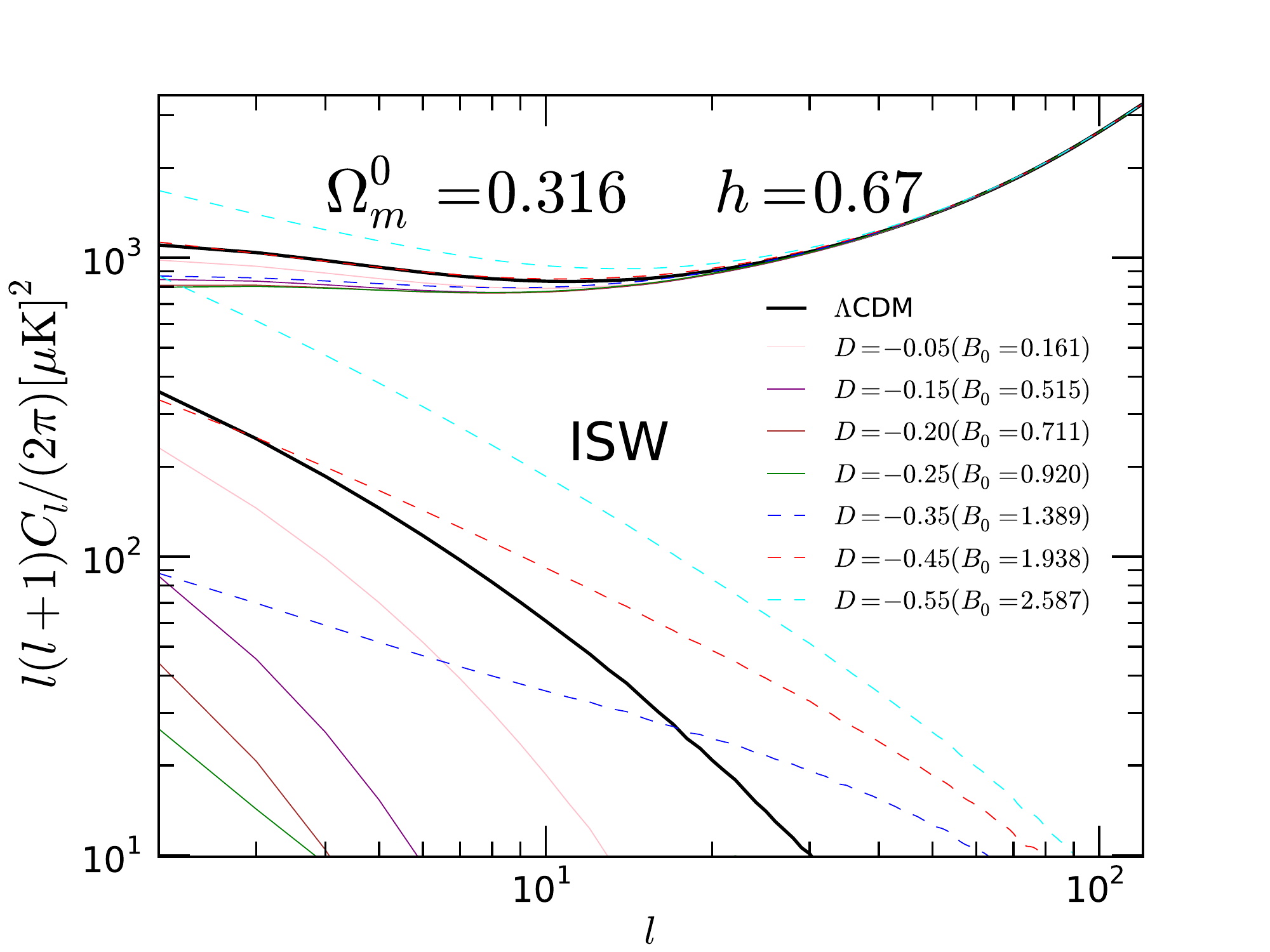}
\caption{The angular power spectrum of the total CMB temperature and the ISW effect at low multipoles. The $f(R)$ gravity will suppress power of the ISW effect(solid lines) and the power reaches its minimal around $D\sim-0.25(B_0\sim0.92)$ then the power turns to grow  as further increasing the value of $|D|$(dashed lines). Around $D\sim-0.45(B_0\sim1.94)$, the power spectrum of the $f(R)$ model almost overlaps with that of the $\Lambda$CDM model. }\label{ISWA}
\end{figure}

Before going further to present our MCMC analysis, we will discuss in this section about how the $f(R)$ gravity impacts the neutrino constraints. The $f(R)$ model studied in this paper actually has rather weak impacts on the early Universe. It only has late-time effects and impacts mainly on the late-time integrated Sachs$-$Wolfe(ISW) effect and the CMB lensing. For the ISW effect, the $f(R)$ gravity will suppress the power of the ISW quadrupole as the parameter $B_0$ which characterizes the $f(R)$ gravity is relatively small~\cite{Song}. As $B_0$ increases, the suppression will reach its maximum and then become reduced.  Further increasing $B_0$, there is a turnaround point above which the suppression will turn into excess, which increases the power of the ISW quadrupole as well as the total quadrupole. In order to better understand this phenomenon, in Fig.\ref{ISWA} we plot the total temperature angular power spectra and the ISW spectra as well, which are calculated by
\begin{equation}
C_l^{\rm ISW}=4\pi \int \frac{dk}{k}\mathcal{P}_{\chi}|\Delta_l^{ISW}(k,\eta_0)|^2\quad,
\end{equation}
where
\begin{equation}
\Delta_l^{ISW}=-2\int_{\eta_i}^{\eta_0}d\eta j_l(k[\eta_0-\eta])e^{-\varepsilon}\left[\frac{d\Phi_{-}}{d\eta}\right]\quad.
\end{equation}
$\mathcal{P}_{\chi}$ is the primordial power spectrum, $j_l(x)$ is the spherical Bessel function, and $\varepsilon$ is the optical depth between $\eta$ and the present. The potential $\Phi_{-}$ which accounts for the ISW effect, is defined by
\begin{equation}
2\Phi_{-}=\Phi-\Psi=-\frac{1}{k}\frac{d\sigma}{d\eta}-\eta_T\quad,\label{len_pot}
\end{equation}
and its derivative with respect to the conformal time $\eta$ is given by
\begin{equation}
2\left[\frac{d\Phi_{-}}{d\eta}\right]=\frac{d\Phi}{d\eta}-\frac{d\Psi}{d\eta}=-\left(\frac{1}{k}\frac{d^2\sigma}{d\eta^2}+\frac{k\sigma}{3}-\frac{k\mathcal{Z}}{3}\right)\quad,
\end{equation}
where $\Phi$ and $\Psi$  in the above equation are the Bardeen potentials\cite{Bardeen} and $\sigma$, $\mathcal{Z}$, $\eta_T$ are the perturbation quantities in the synchronous gauge. We present the equivalent expressions in the synchronous gauge for $\Phi_{-}$ here because the CAMB code is based on the synchronous gauge.

\begin{figure}
  \includegraphics[width=3in,height=2.8in]{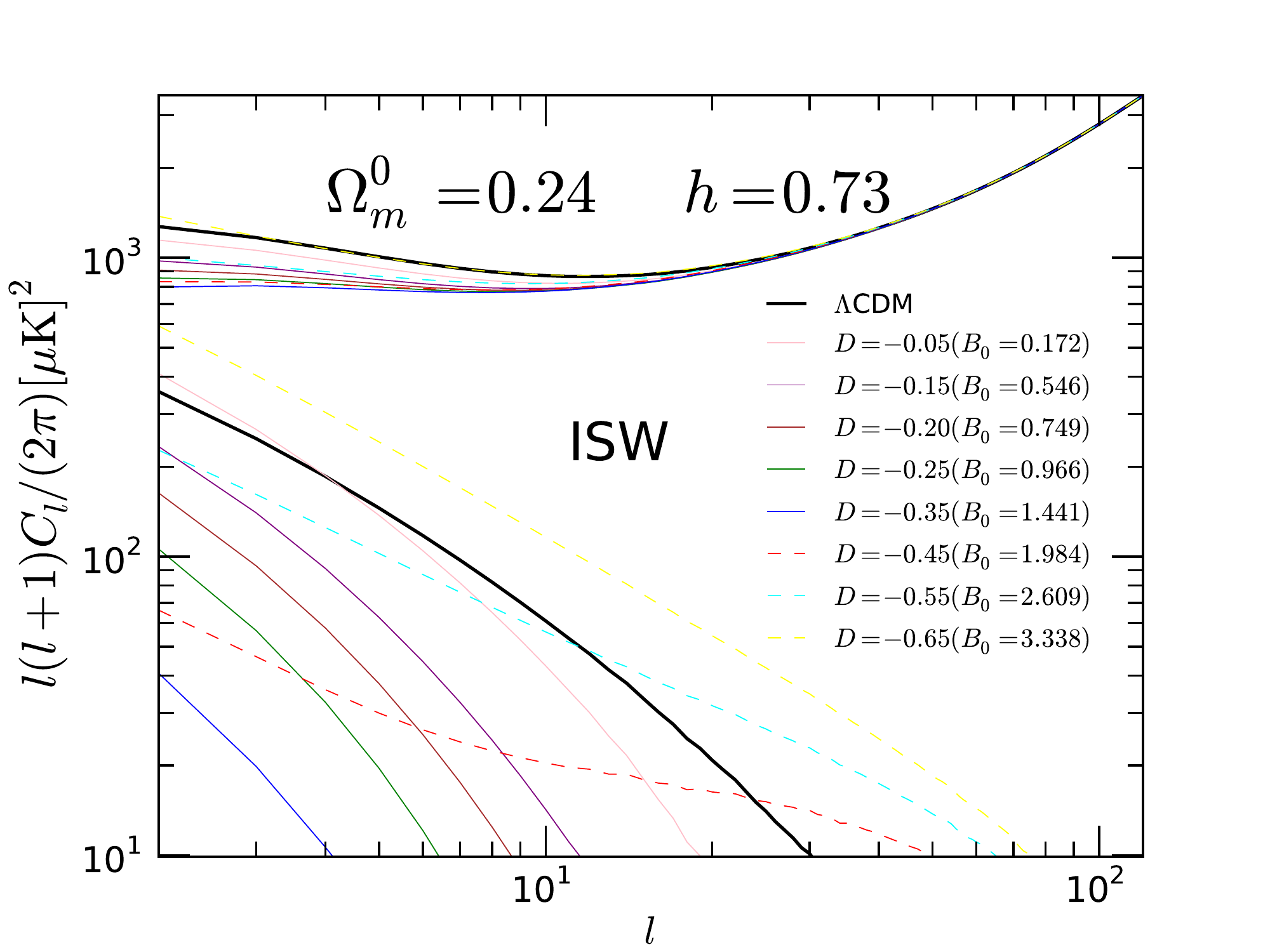}
\caption{Similar to Fig.~\ref{ISWA} but with different cosmological parameters. The suppression (solid lines) reaches its maximal around $D\sim-0.37(B_0\sim1.5)$ and then around $D\sim-0.60(B_0\sim 3)$ the power spectrum goes back (dashed lines) to that of the $\Lambda$CDM model.}\label{ISWB}
\end{figure}

For illustrative purposes, we take the cosmological parameters for the fiducial model as the best-fitted values of the $\Lambda$CDM model as reported by the Planck team $\Omega_b^0=0.049,\Omega_c=0.267,\Omega_{\Lambda}=0.684, h=0.6711, n_s=0.962,10^{9}A_s=2.215,\tau=0.0925$\cite{planck}. From Fig.\ref{ISWA}, we can see that the suppression of the ISW power spectra reaches its maximal around $D\sim-0.25(B_0\sim0.92)$ then the power turns to grow from its minimal as further increasing the value of $|D|$. Around $D\sim-0.45(B_0\sim1.94)$, the power spectra of the $f(R)$ model go back to being similar to that of the $\Lambda$CDM model. The $f(R)$ gravity and the $\Lambda$CDM model give almost the same temperature angular power spectrum at this point. However, the value of $B_0$ for this point depends on the cosmological parameters of the fiducial model. To show this, in Fig.\ref{ISWB}, we plot the power spectra of the model with different values of $\Omega_m=0.24$ and $h=0.73$ which are the same as those used in Ref.~\cite{Song} and keep the other cosmological parameters unchanged. We find that around $D\sim-0.37(B_0\sim1.5)$ , the suppression reaches its maximal and around $D\sim-0.60(B_0\sim 3)$, the power spectra go back to being similar to that of the $\Lambda$CDM model. Our results are actually well consistent with Ref.~\cite{Song}, if we take the same values of the cosmological parameters.

\begin{figure}
\includegraphics[width=3in,height=2.8in]{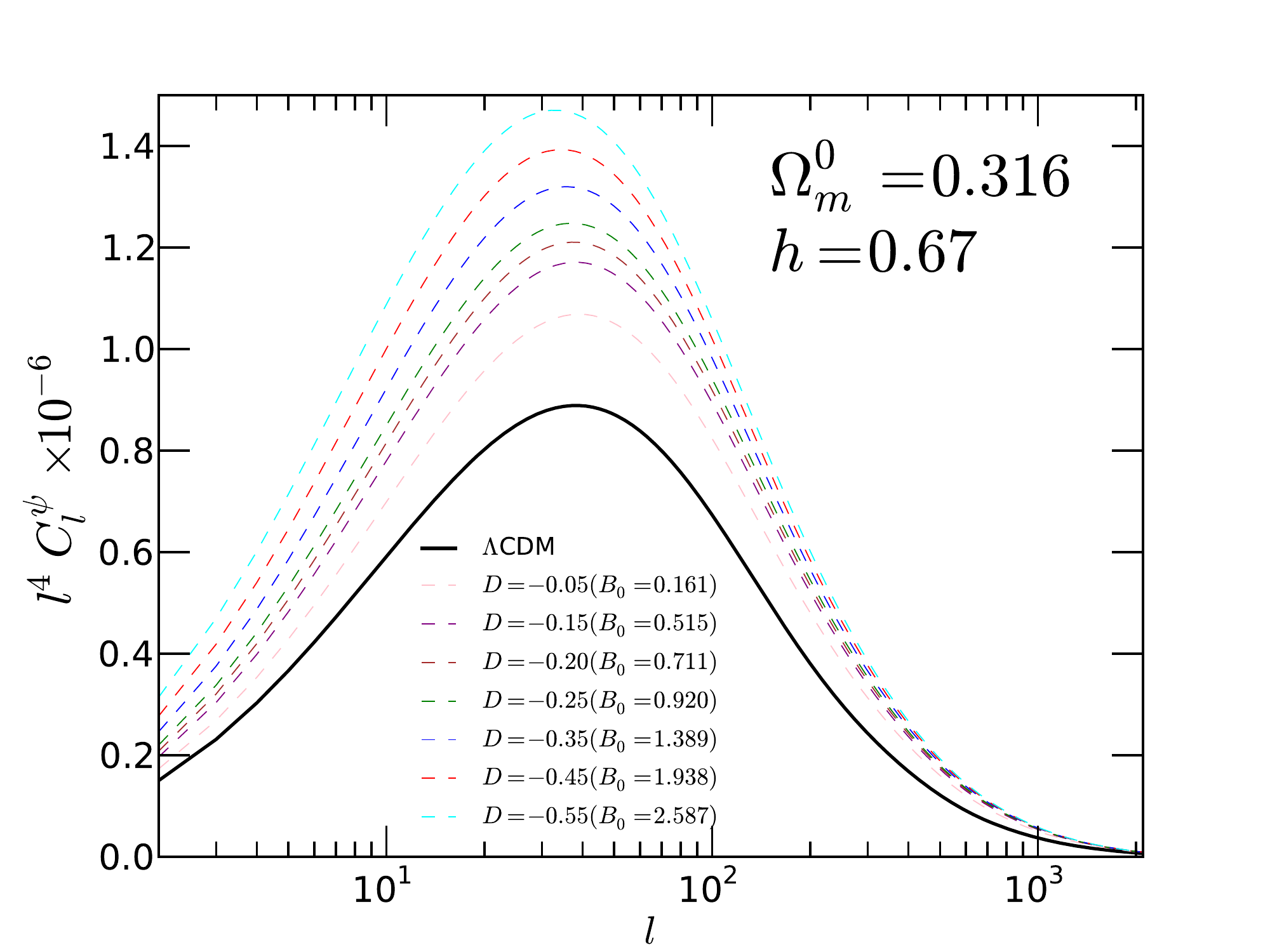}
\caption{The impact of the $f(R)$ gravity on the angular power spectrum of the lensing potential. }\label{lensing}
\end{figure}

In Fig.~\ref{lensing}, we show the angular power spectra of the lensing potential $\psi\equiv-\Phi_{-}$ for a few representative values of $D$. From Fig.~\ref{lensing}, we can see that, contrary to the ISW effect, $f(R)$ gravity always enhances the power of the lensing potential. The larger the value of $|D|$, or equivalently, of $B_0$, the more enhancement in the power spectrum of the potential. We should remark here that in the original CAMB code, $\psi$ is calculated by using an approximation. However, this approximation does not apply to the $f(R)$ gravity. We need to use the exact expression of Eq.(\ref{len_pot}) to calculate $\psi$ instead. Then we follow the standard routine in the CAMB code to calculate $C_{l}^{\psi}$. The detailed derivations of  $C_{l}^{\psi}$ can be found in Ref.~\cite{lensing_review}.

\begin{figure}
\includegraphics[width=3in,height=2.8in]{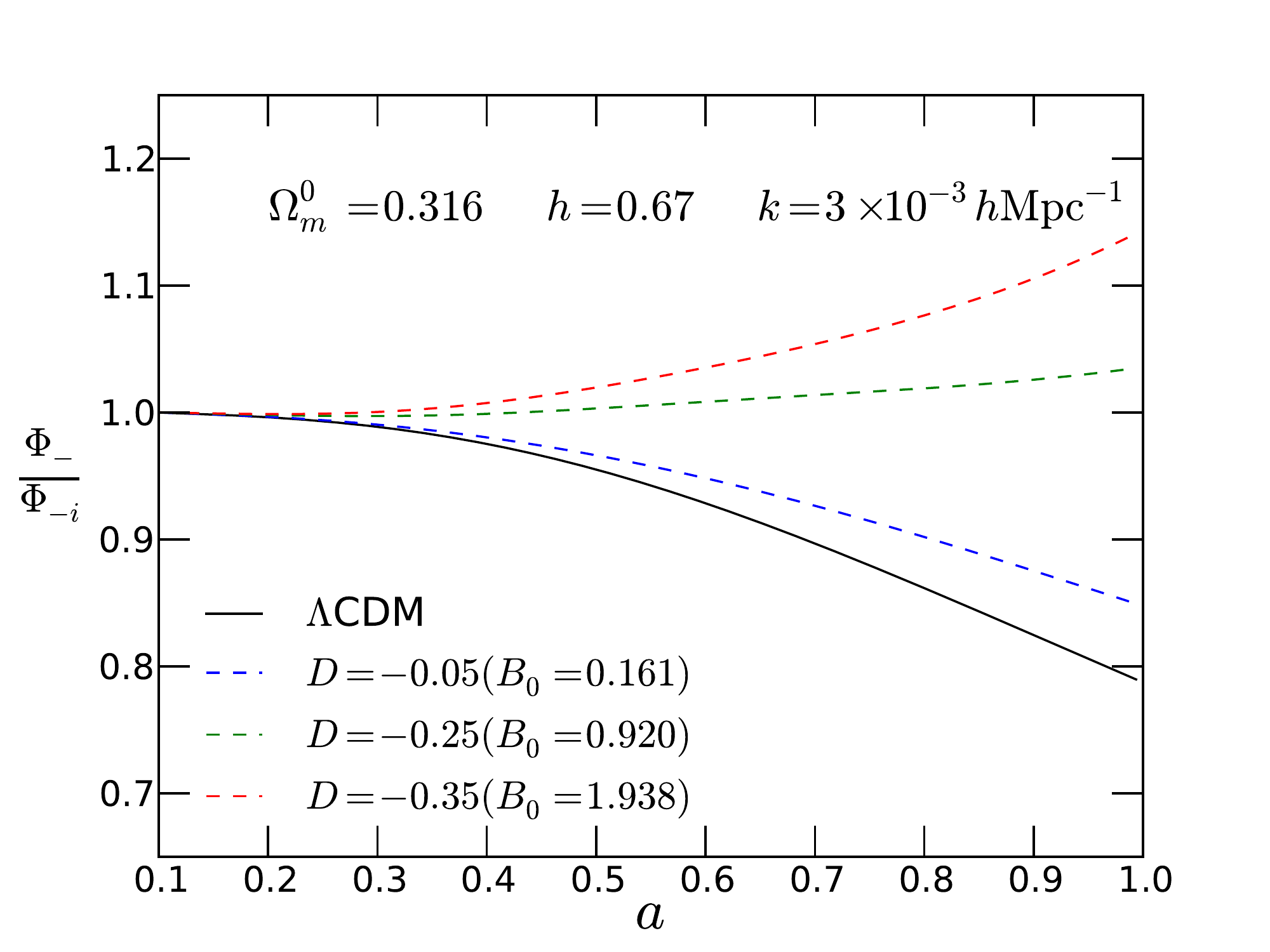}
\caption{Evolution of metric fluctuations $\Phi_{-}$ for the $\Lambda$CDM model and a few representative values of $D$ in the $f(R)$ models. $\Phi_{-i}$ is the value of the potential in the $\Lambda$CDM model at $a=0.03$. The potential $\Phi_{-}$ is always enhanced in the $f(R)$ gravity.
 }\label{Potential}
\end{figure}

The phenomenon as described above of the impact of the $f(R)$ gravity on the ISW effect (e.g. Fig.\ref{ISWA},Fig.\ref{ISWB}) and the CMB lensing (e.g. Fig.\ref{lensing}) can be explained by the evolution of the metric potential $\Phi_{-}$\cite{Song}. In Fig.~\ref{Potential}, we show the value of $\Phi_{-}/\Phi_{-i}$ with respect to the scale factor $a$. We choose the wave number as $k=3\times 10^{-3} h{\rm Mpc^{-1}}$ from which the power of the quadrupole mainly arises\cite{Song}. $\Phi_{-}$ is calculated by Eq.(\ref{len_pot}) and $\Phi_{-i}$ is the value of the potential in the $\Lambda$CDM model at $a=0.03$. As is well-known, the ISW effect is driven by the evolution of the potential $\Phi_{-}$, which depends on the relative difference of the potential $\Phi_{-}$ at the initial time $\Phi_{-i}$ and the present time $\Phi_{-0}$. From Fig.~\ref{Potential}, we can see that the gravitational potential $\Phi_{-}$ always decays in the $\Lambda$CDM model at late times of the Universe. However, in the $f(R)$ gravity, the potential  will be enhanced against such decay due to the existence of the extra scalar field $\delta F$. $\Phi_{-}$ in the $f(R)$ gravity will decay less than that in the $\Lambda$CDM model when the value of $B_0$ is relatively small (e.g. $B_0=0.161$). Then, for a certain value of $B_0$ (e.g.$B_0\sim0.920$), $\Phi_{-0}$ at present will be comparable to $\Phi_{-i}$ at early times. The ISW effect is canceled out at this point. For large enough $B_0$ (e.g. $B_0=1.938$), the potential at present will overwhelm the potential at early times $\Phi_{-0}>\Phi_{-i}$ and the ISW effect in the $f(R)$ gravity will change its sign. However, the amplitude of the ISW effect increases with $B_0$ as $B_0$ becomes much larger. This explains what we observed in Fig.~\ref{ISWA} and Fig.~\ref{ISWB}. For the CMB lensing, we can find that, contrary to the ISW effect, the angular power spectrum of the lensing potential $C_{l}^{\psi}$~\cite{lensing_review} depends on the absolute value of the amplitude of the potential $\Phi_{-}$, which increases monotonously with $B_0$ as shown in Fig.~\ref{Potential}.  It is the case, therefore, that the larger the value of $B_0$, the larger the power of $C_{l}^{\psi}$.

\begin{figure}
\includegraphics[width=3in,height=2.8in]{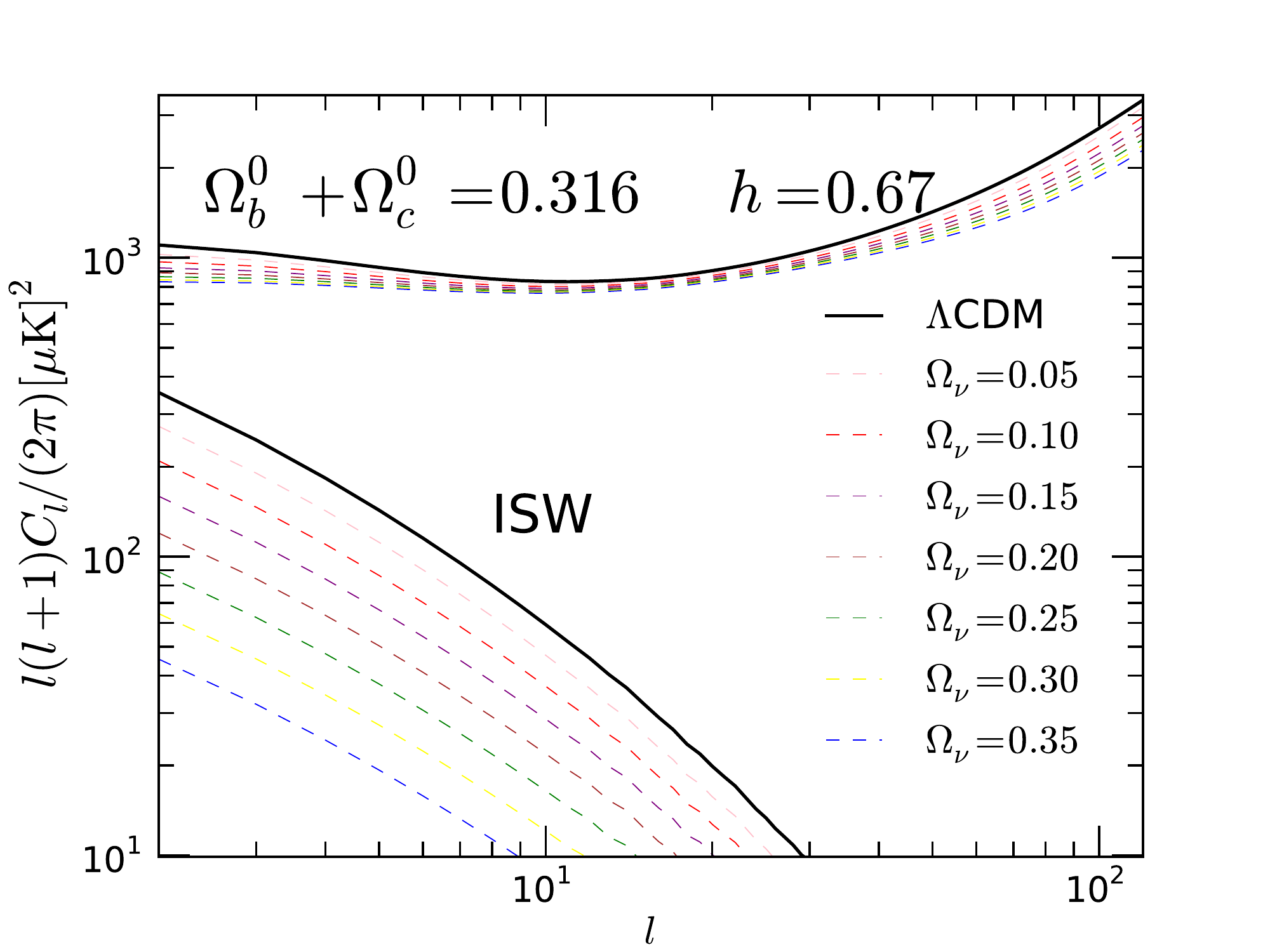}
\caption{The impact of massive neutrinos on the temperature angular power spectrum and ISW effect. }\label{ISWmnu}
\end{figure}

\begin{figure}
\includegraphics[width=3in,height=2.8in]{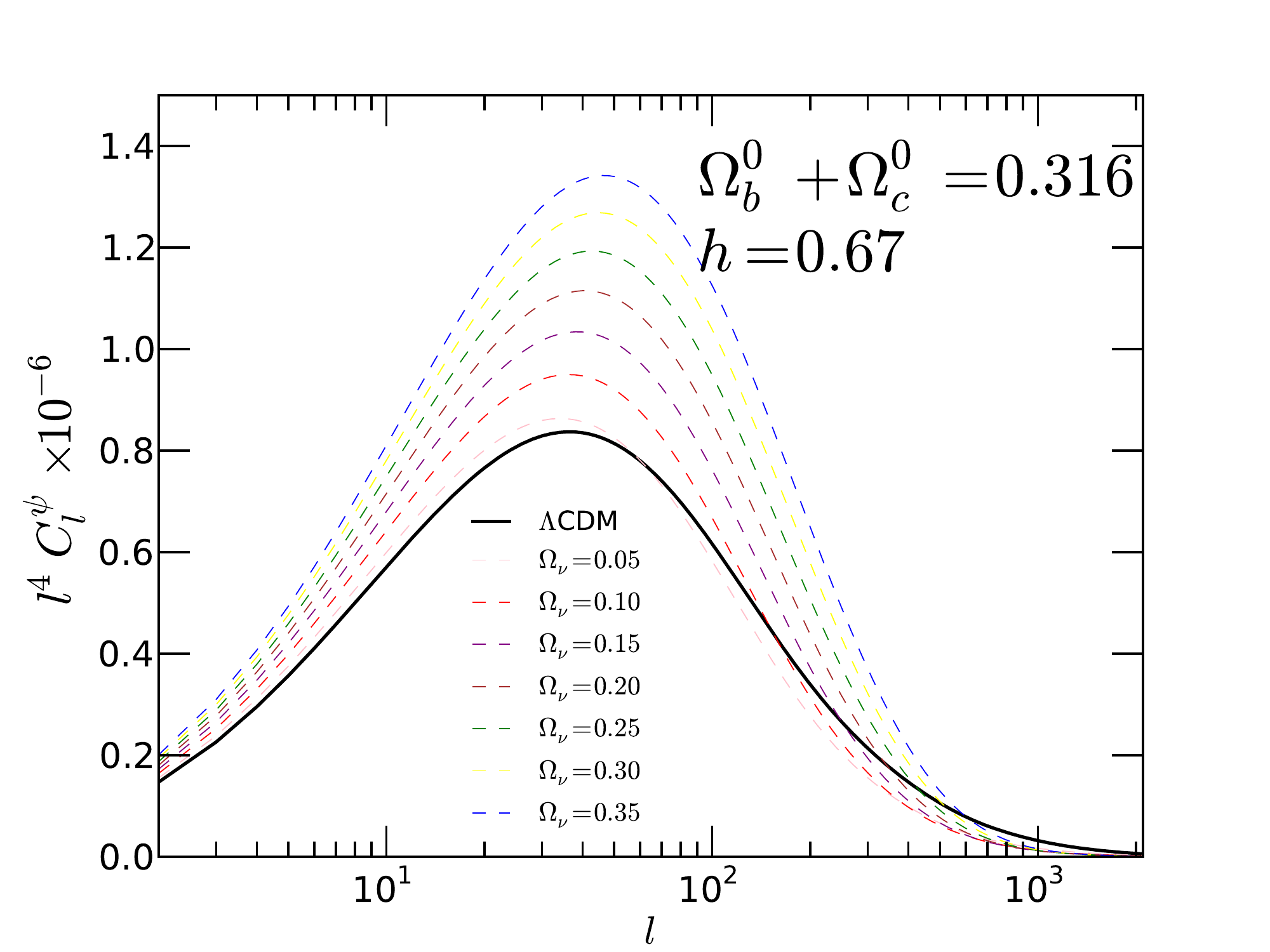}
\caption{The impact of massive neutrinos on the angular power spectrum of the lensing potential. }\label{lensingmnu}
\end{figure}

On the other hand, neutrinos with mass heavier than a few ${\rm eV}$ will become nonrelativistic before the recombination, triggering significant impact on the CMB anisotropy spectrum. However, this situation is strongly disfavoured by current observational bounds even in the case of $f(R)$ gravity as we shall see later. Therefore, we will not discuss this case here. Neutrinos with a mass ranging from $10^{-3}$ eV to $1{\rm eV}$ will be relativistic at the time of matter-radiation equality and will be nonrelativistic today, which can potentially impact the CMB in three ways(see \cite{Julien} for reviews). The massive neutrino can shift the redshift of equality which affects the position and amplitude of the peaks; it can also change the angular diameter distance to the last scattering surface which affects the overall position of CMB spectrum features; the massive neutrino can affect the late time ISW effect as well.  We will focus on the ISW effect in this work. In Fig.~\ref{ISWmnu}, we plot the total angular power spectrum and ISW effect for a few representative values of the density of the massive neutrinos $\Omega_{\nu}$. We can see that the massive neutrinos will suppress the power of the ISW effect and the power of the total power spectrum. We also plot the impact of the massive neutrinos on the angular power spectrum of the lensing potential in Fig.~\ref{lensingmnu}. We can see that the massive neutrinos will always enhance the power of the lensing potentials.

From the above analysis, we can see that with the cosmological parameters of the fiducial model around $\Omega_m\sim 0.32$,$h\sim0.67$, which is favored by the Planck results \cite{planck}, if $B_0<0.92$, the impact of $f(R)$ gravity on the ISW effect and the CMB lensing is degenerate with the impact of the massive neutrinos. Moreover, for $f(R)$ models with $B_0>0.92$, the impact of $f(R)$ gravity on the ISW effect could partially compensate the effect of massive neutrinos since $f(R)$ gravity enhances the power as $B_0$ grows if $B_0>0.92$. This compensation would further boost the degeneracy between $B_0$ and $\sum m_{\nu}$ as we shall see later.

\section{Current Observational Data\label{Data}}
In this work, we adopt the CMB data from the Planck satellite\cite{planck}, as well as the high-$l$ data from the Atacama Cosmology Telescope(ACT)\cite{ACT} and the South Pole Telescope(SPT)\cite{SPT}. For the Planck data, we use the likelihood code provided by the Planck team, which includes the high-multipoles $l>50$ likelihood following the \texttt{CamSpec} methodology and the low-multipoles ($2<l< 49$) likelihood based on a Blackwell-Rao estimator applied to Gibbs samples computed by the \texttt{Commander} algorithm. For the high-$l$ data, we include the ACT $148\times148$ spectra for $l\geq1000$, and the ACT $148\times218$ and $218\times218$ spectra for $l\geq1500$. For SPT data, we only use the high multipoles with $l>2000$. In our analysis, the WMAP polarization data will be used along with Planck temperature data.

For comparison, we also present the results obtained from WMAP nine-year data in this work. The likelihood code\cite{wmap9} contains both temperature and polarization data.
The temperature data include the CMB anisotropies on scales $2 \leq l \leq 1200$;the polarization data contain TE/EE/BB power spectra on scales $(2 \leq l \leq 23)$
and TE power spectra on scales $(24 \leq l \leq 800)$.

In addition to the CMB data, we also add the measurement on the distance indicator from the baryon acoustic oscillations(BAO) surveys.
BAO surveys measure the distance ratio between $r_s(z_{\rm drag})$ and $D_v(z)$
\begin{equation}
d_z=\frac{r_s(z_{\rm drag})}{D_v(z)}\quad,
\end{equation}
where $r_s(z_{\rm drag})$ is the comoving sound horizon at the baryon drag epoch, which is defined by
\begin{equation}
r_s(z)=\int_0^{\eta(z)}\frac{d\eta}{\sqrt{3(1+R)}}\quad,
\end{equation}
where $\eta$ is the conformal time and $R\equiv 3\rho_b/(4\rho_{r})$. The drag redshift $z_{\rm drag}$
 indicates the epoch for which the Compton drag balances the gravitational force, which happens at $g_d\sim 1$, where
\begin{equation}
g_d(\eta)=\int_{\eta_0}^{\eta}\dot{g}d\eta/R\quad,
\end{equation}
with $\dot{g}=-a n_e \sigma_T$(where $n_e$ is the density of free electrons and $\sigma_T$ is the Thomson cross section). $z_{\rm drag}$ is defined by $g_d(\eta(z_{\rm drag}))=1$.
The quantity $D_{v}(z)$ is a combination of the angular diameter distance $D_{A}(z)$ and the Hubble parameter $H(z)$.
\begin{equation}
D_{v}(z)=\left[(1+z)^2D_{A}^2(z)\frac{cz}{H(z)}\right]^{1/3}\quad.
\end{equation}
\begin{figure}
\includegraphics[width=3in,height=2.8in]{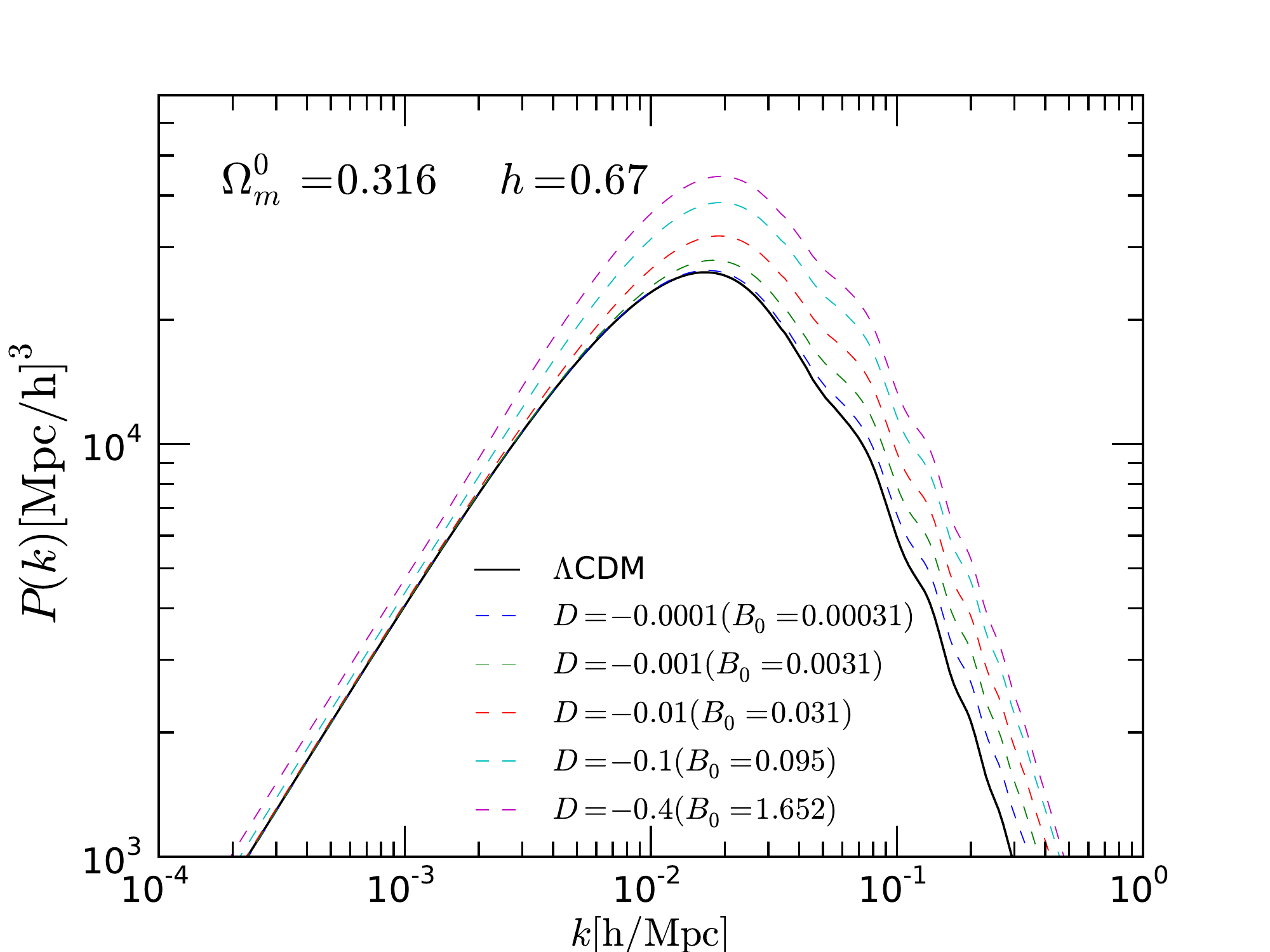}
\caption{Linear matter power spectrum for a few representative values of $D$ at redshift $z=0$. It is clear that the scale-dependent growth history changes not only the amplitude but also the shape of the matter power spectra.}\label{pk}
\end{figure}
\begin{figure}
\includegraphics[width=3in,height=2.8in]{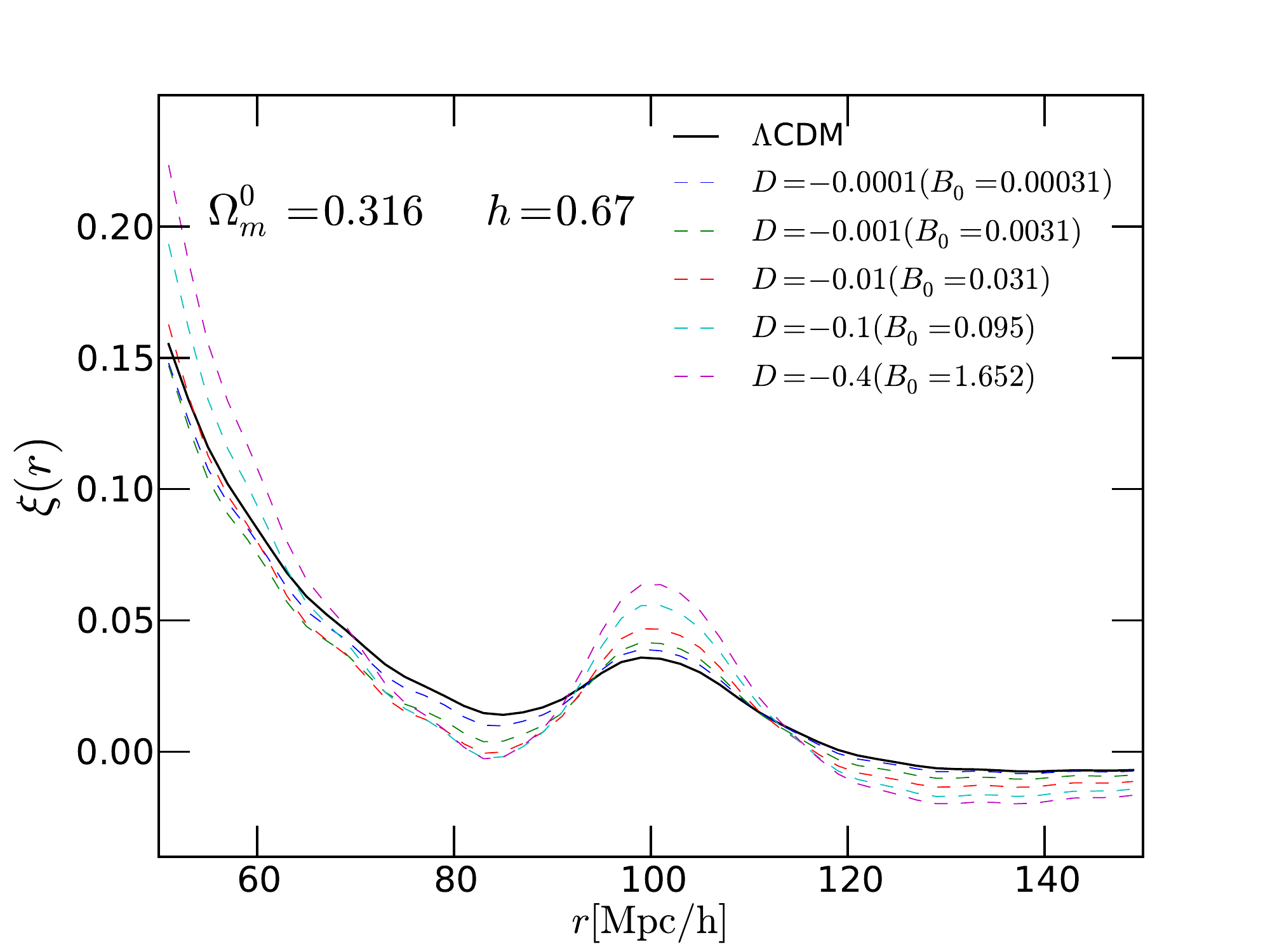}
\caption{The 2-point correlation function in real-space. Although the shape is sensitive to the value of $D$, the BAO scale does not change in this family of $f(R)$ models. }\label{cor}
\end{figure}
Although the $f(R)$ model studied in this work exhibits strong
scale-dependent growth history even in the linear regime(see Fig~\ref{pk}), which changes not only the amplitude but also the shape of the matter power spectra, in real space the scale of the BAO peak in the two-point correlation function of the density field does not change for this family of $f(R)$ models:
\begin{equation}
\xi(r)=\frac{1}{2\pi^2}\int dk k^2P_L(k)\frac{\sin(kr)}{kr}\quad.
\end{equation}
From Fig~\ref{cor}, we can see that the BAO scales do not shift in this family of $f(R)$ models. The locations of the BAO peaks in the $f(R)$ models relative to that in the $\Lambda$CDM model shift no more than $\pm 1.5 {\rm Mpc/h}$, which is mainly subject to the numerical errors.
In this paper, we therefore can safely adopt the BAO data.
We follow the Planck analysis \cite{planck} and use the BAO measurements from four different redshift surveys:$z=0.57$ from the BOSS DR9 measurement \cite{BAO1};
$z=0.1$ from the 6dF Galaxy Survey measurement \cite{BAO2};
$z=0.44,0.60$ and $0.73$ from the WiggleZ measurement\cite{BAO3};
$z=0.2$ and $z=0.35$ from the SDSS DR7 measurement\cite{BAO4}.
\section{Numerical results\label{Num}}
In this section, we explore the cosmological parameter space in our $f(R)$ model using the Markov chain Monte Carlo analysis. Our analysis is based on the public available code COSMOMC \cite{mcmc} as well as a modified version of the CAMB code which solves the full linear perturbation equations in the $f(R)$ gravity~\cite{frlinear}. The parameter space of our model is
\begin{equation}
P=(\Omega_bh^2,\Omega_ch^2,100\theta_{\rm MC},\ln[10^{10}A_s],n_s,\tau,\sum m_{\nu},N_{\rm eff},D)\quad,
\end{equation}
where $\Omega_bh^2$ and $\Omega_c h^2$ are the physical baryon and cold dark matter energy densities respectively, $100\theta_{\rm MC}$ is the angular size of the acoustic horizon, $A_s$ is the amplitude of the primordial curvature perturbation, $n_s$ is the scalar spectrum power-law index, $\tau$ is the optical depth due to reionization, $\sum m_{\nu}$ is the sum of the neutrino mass in eV, $N_{\rm eff}$ is the effective number of neutrinolike relativistic degrees of freedom and $D$ is the parameter which characterizes the $f(R)$ gravity. We will sample the parameter $D$ directly in our work and treat $B_0$ as a derived parameter.  The priors for the cosmological parameters are listed in Table~\ref{priors}.
\begin{table*}
\caption{Uniform priors for the cosmological parameters \label{priors}}
\begin{tabular}{c}
\hline
$0.005<\Omega_bh^2<0.1$ \\
$0.001<\Omega_ch^2<0.99$\\
$0.5<100\theta_{\rm MC}<10.0$ \\
$0.01<\tau<0.8$  \\
$0.9<n_s<1.1$  \\
$2.7<\rm{ln}[10^{10}As]<4.0$ \\
$-1.2<D<0$ \\
$0<\sum m_{\nu}<5$ \\
$0.05<N_{\rm eff}<10.0$ \\
\hline
\end{tabular}
\end{table*}

In this work, we will pay particular attention to the neutrino properties. We will fix $N_{\rm eff}=3.046$ to constrain the total mass of neutrinos $\sum m_\nu$ and, in turn, fix $\sum m_\nu=0.06[{\rm eV}]$ to constrain the effective number of neutrino species $N_{\rm eff}$. Finally, we will constrain $N_{\rm eff}$ and $\sum m_\nu$ simultaneously.

\subsection{Constraints on the total mass of active neutrinos}
\begin{figure*}
\includegraphics[width=5in,height=4in]{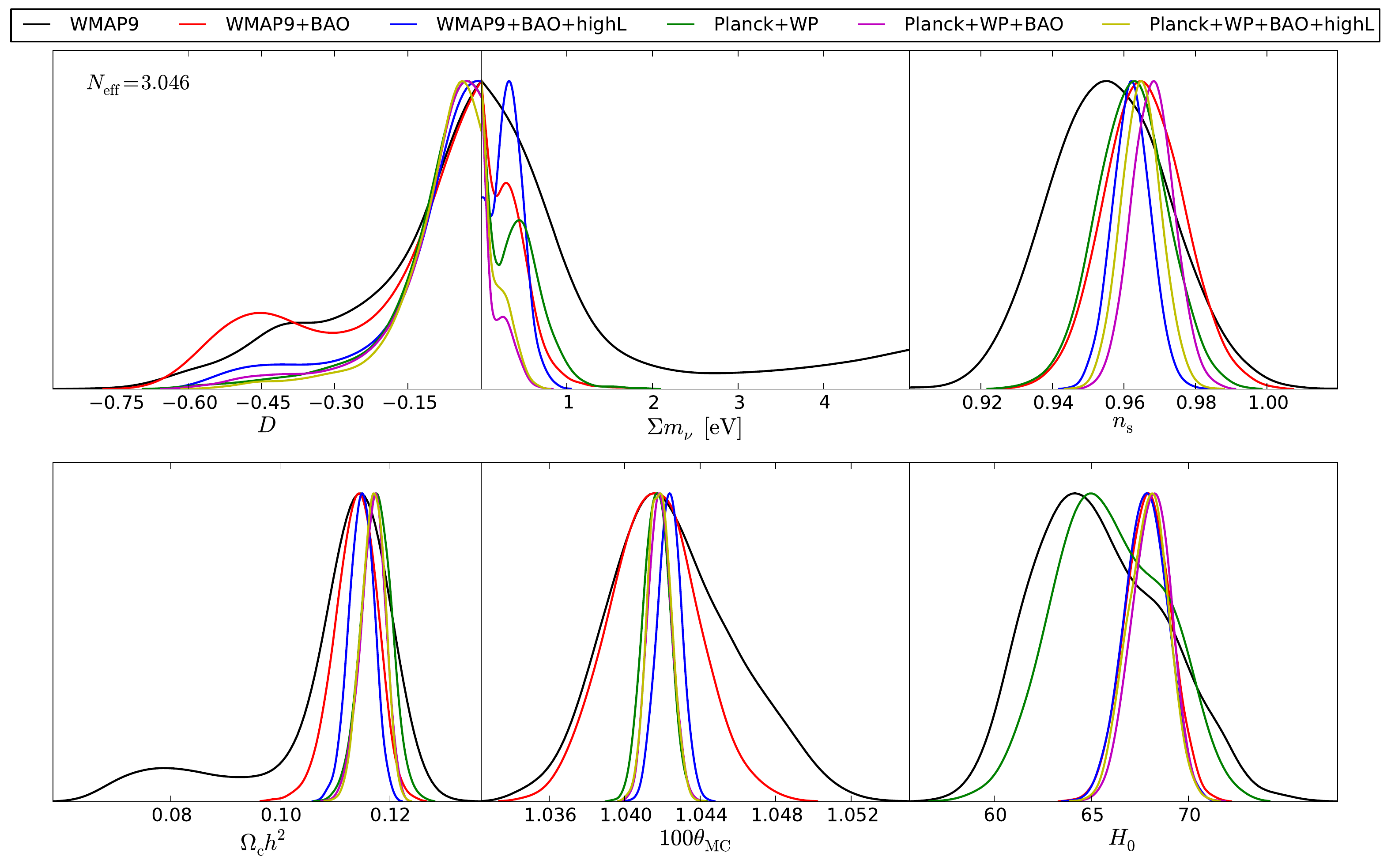}
\caption{One-dimensional marginalized likelihood for the total neutrino mass $\sum m_{\nu}$ as well as other cosmological parameters $D,\quad n_s,\quad\Omega_ch^2,\quad 100\theta{\rm MC},\quad H_0$. In these $f(R)$ models, we set $N_{\rm eff}=3.046$.\label{numass1D}}
\end{figure*}

In this subsection, we report the constraints on the total mass of active neutrinos $\sum m_{\nu}$ assuming $N_{\rm eff}=3.046$. The numerical results are shown in Table \ref{numass}. In Fig.\ref{numass1D}, we show the one-dimensional marginalized likelihood for the total neutrino mass $\sum m_{\nu}$ as well as other cosmological parameters $D,\quad n_s,\quad\Omega_ch^2,\quad 100\theta{\rm MC},\quad H_0$. We start by presenting the results obtained from the data combinations associated with WMAP nine-year data . From Table\ref{numass}, we can find that WMAP nine-year data along place very poor constraints on $\sum m_{\nu}$ ,$\Omega_ch^2$ and $H_0$.  $\sum m_{\nu}$ remains almost unconstrained and the $2\sigma(95\%{\rm C.L.})$  range of marginalized likelihood for $\sum m_{\nu}$ almost spans the whole range as our priors listed in Table~\ref{priors}. However, if we add the BAO data, the constraint can be improved significantly, because the BAO data can improve the constraint on $H_0$ and breaks the degeneracy between $H_0$ and $\sum m_{\nu}$. The combination of WMAP+BAO gives $$\sum m_{\nu}<0.802{\rm eV}(95\%{\rm C.L.};{\rm WMAP9+BAO})\quad.$$
However, adding the BAO data does not improve the constraint on $f(R)$ gravity. We find $ D<0.542( B_0<2.54)(95\%{\rm C.L.};{\rm WMAP+BAO})$ which is even slightly larger than the constraints obtained from WMAP data alone $ D<0.518(B_0<2.37)(95\%{\rm C.L.};{\rm WMAP})$.
Adding the high-$l$ measurement from the CMB can further improve the constraint on $\sum m_{\nu}$ because the WMAP data do not have enough accuracy on the high-$l$ angular power spectra. The combination of WMAP9+BAO+highL places the constraint at $$\sum m_{\nu}<0.608{\rm eV}(95\%{\rm C.L.};{\rm WMAP9+BAO+highL})\quad.$$

\begin{figure}
\includegraphics[width=3in,height=2.8in]{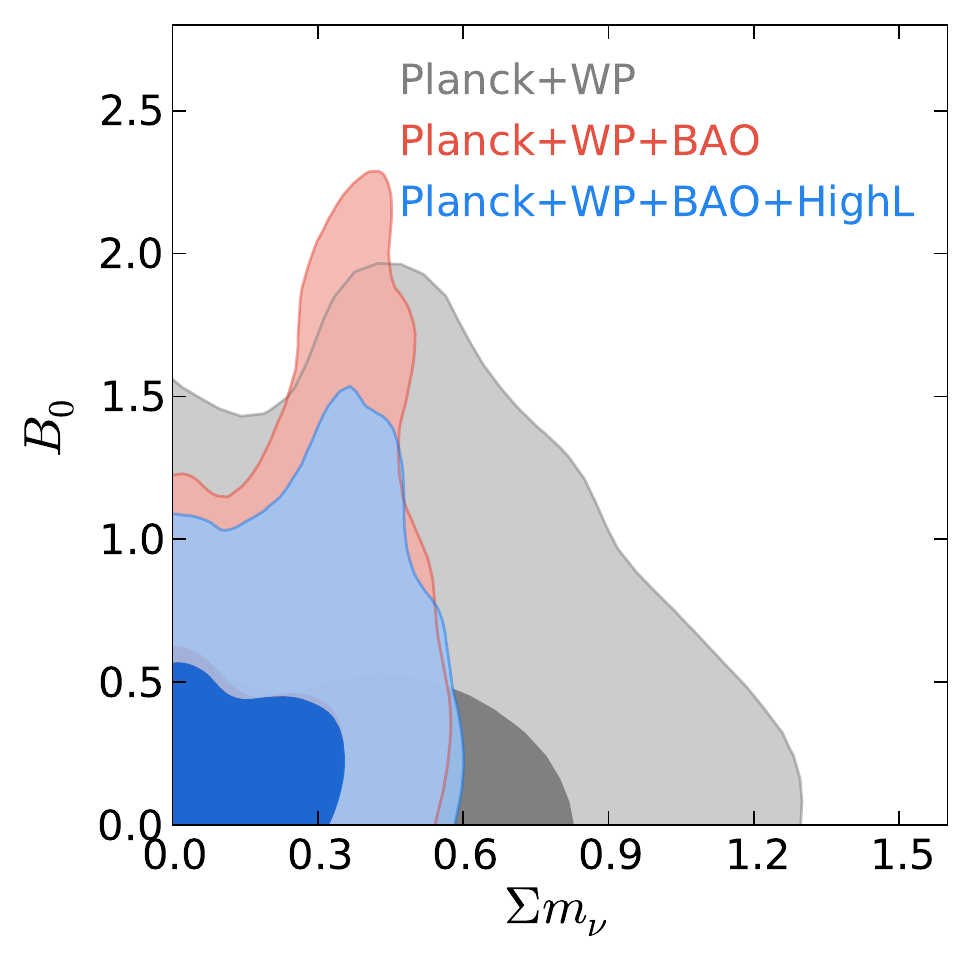}
\caption{Marginalized two-dimensional likelihood ($1, 2\sigma$ contours) constraints on $B_0$ and $\sum m_{\nu}$. There are degeneracies between these two parameters. When $B_0>1$, there are tails in the contours, which means the degeneracy sharpens here. This is because the impact of $f(R)$ gravity on the ISW effect could be partially compensated by the impact of massive neutrinos if $B_0>1$. }\label{contour}
\end{figure}

Compared with the constraints associated with WMAP data, Planck data  show more robust constraints on $\sum m_{\nu}$ as well as the $f(R)$ gravity. Although the Planck data alone in combination with WMAP polarization(WP) data only place very weak constraints on the total neutrino mass, $$ \sum m_{\nu}<0.928{\rm eV}(95\%{\rm C.L.};{\rm Planck+WP})\quad,$$ they put tighter constraints on the $f(R)$ gravity $D <0.346(B_0<1.36)$(95\%{\rm C.L.}) due to fact that $f(R)$ gravity produces the quadrupole suppression on the temperature angular power spectra\cite{Song} and the Planck data have a more accurate measurement on the large-scale ($2<l<50$) temperature angular power spectra than that of the WMAP data. The data combination Planck+WP, however, can not put a tight constraint on $H_0$, as shown in Fig.\ref{numass1D}. Planck+WP therefore gives very poor constraint on $\sum m_{\nu}$ due to the degeneracy between $H_0$ and $\sum m_{\nu}$. Therefore, it can be expected that adding BAO data can improve the constraints significantly. We find
$$\sum m_{\nu}<0.463{\rm eV}(95\%{\rm C.L.};{\rm Planck+WP+BAO})\quad,$$ with $ |D|<0.379(B_0<1.54)$(95\%{\rm C.L.}). The constraint on $\sum m_{\nu}$ has been improved by almost $50\%$ by adding the BAO data. On the other hand, we find that the high-$l$ data do not show a significant improvement on the constraint of $\sum m_{\nu}$ but slightly improve on the constraint of $f(R)$ gravity due to the tighter constraint on $\Omega_ch^2$(see Table ~\ref{numass}). We find
$$\sum m_{\nu}<0.462{\rm eV}(95\%{\rm C.L.};{\rm Planck+WP+BAO+highL})$$ and $ |D|<0.298(B_0<1.14)$(95\%{\rm C.L.}). In order to show the degeneracy between $B_0$ and $\sum m_{\nu}$. We plot the Marginalized two-dimensional likelihood ($1, 2\sigma$ contours) constraints on $B_0$ and $\sum m_{\nu}$ in Fig~\ref{contour}. We can see that when $B_0>1$, there are tails in the contours, which means the degeneracy sharpens here. This is because the impact of $f(R)$ gravity on the ISW effect could partially be compensated by the massive neutrinos if $B_0>1$ as discussed previously.
\begin{table*}
\caption{Cosmological parameter values for the $f(R)$ models with $N_{\rm eff}=3.046$. $B_0$ is a derived parameter. \label{numass}}
\begin{tabular}{|c||c|c|c|c|c|c|}
\hline
Parameters & WMAP9 & WMAP9+BAO & WMAP9+BAO+highL &Planck+WP & Planck+WP+BAO& Planck+WP+BAO+highL\\
\hline
$\Omega_bh^2$ &$0.02288^{+0.00072}_{-0.00072}$ &$0.02278^{+0.00050}_{-0.00050}$&$0.02273^{+0.00026}_{-0.00026}$&$0.02252^{+0.00035}_{-0.00035}$&$0.02264^{+0.00032}_{-0.00032}$ &$0.02259^{+0.00029}_{-0.00029}$ \\
\hline
$\Omega_ch^2$ &$0.1090^{+0.0130}_{-0.0130}$ &$0.1138^{+0.0039}_{-0.0039}$&$0.1149^{+0.0023}_{-0.0023}$&$0.1176^{+0.0028}_{-0.0028}$&$0.1170^{+0.0023}_{-0.0023}$&$0.1170^{+0.0021}_{-0.0021}$\\
\hline
$100\theta_{\rm MC}$ & $1.0426^{+0.0033}_{-0.0033}$
& $1.0415^{+0.0023}_{-0.0023}$&$1.0424^{+0.0006}_{-0.0006}$ & $1.0417^{+0.0007}_{-0.0007}$& $1.0419^{+0.0006}_{-0.0006}$&$1.0418^{+0.0006}_{-0.0006}$\\
\hline
$\tau$ &$ 0.0835^{+0.0126}_{-0.0126}$ & $0.0854^{+0.0128}_{-0.0128}$& $0.0813^{+0.0114}_{-0.0114}$& $0.0809^{+0.012}_{-0.012}$& $0.0822^{+0.012}_{-0.012}$&$0.0815^{+0.012}_{-0.012}$ \\
\hline
$n_s$ & $0.9562^{+0.0163}_{-0.0163}$ & $0.9656^{+0.0111}_{-0.0111}$&$0.9621^{+0.0053}_{-0.0053}$& $0.9621^{+0.0099}_{-0.0099}$& $0.9682^{+0.0060}_{-0.0060}$&$0.9648^{+0.0055}_{-0.0055}$ \\
\hline
$\rm{ln}[10^{10}As]$ & $3.074^{+0.027}_{-0.027}$ &$3.077^{+0.029}_{-0.029}$ &$3.060^{+0.021}_{-0.021}$&$3.063^{+0.023}_{-0.023}$&$3.065^{+0.025}_{-0.025}$& $3.062^{+0.022}_{-0.022}$\\
\hline
$|D|$ & $ <0.518$(95\%{\rm C.L.}) & $ <0.542$(95\%{\rm C.L.})&$ <0.452$(95\%{\rm C.L.})& $ <0.346$(95\%{\rm C.L.})& $ <0.379$(95\%{\rm C.L.})& $ <0.298$(95\%{\rm C.L.})\\
($B_0$) & $ <2.37$(95\%{\rm C.L.}) & $ < 2.54$(95\%{\rm C.L.})&$ < 1.99$(95\%{\rm C.L.})& $ <1.36$(95\%{\rm C.L.})& $ <1.54$(95\%{\rm C.L.})& $ <1.14$(95\%{\rm C.L.})\\
\hline
$\sum m_{\nu}[{\rm eV}]$ & $ <5$(95\%{\rm C.L.}) & $ <0.802$(95\%{\rm C.L.})&$ <0.608$(95\%{\rm C.L.}) &$ <0.928$(95\%{\rm C.L.})& $ <0.463$(95\%{\rm C.L.})&$ <0.462$(95\%{\rm C.L.})\\
\hline
\end{tabular}
\end{table*}

\subsection{Constraints on $N_{\rm eff}$}
\begin{figure*}
\includegraphics[width=5in,height=4in]{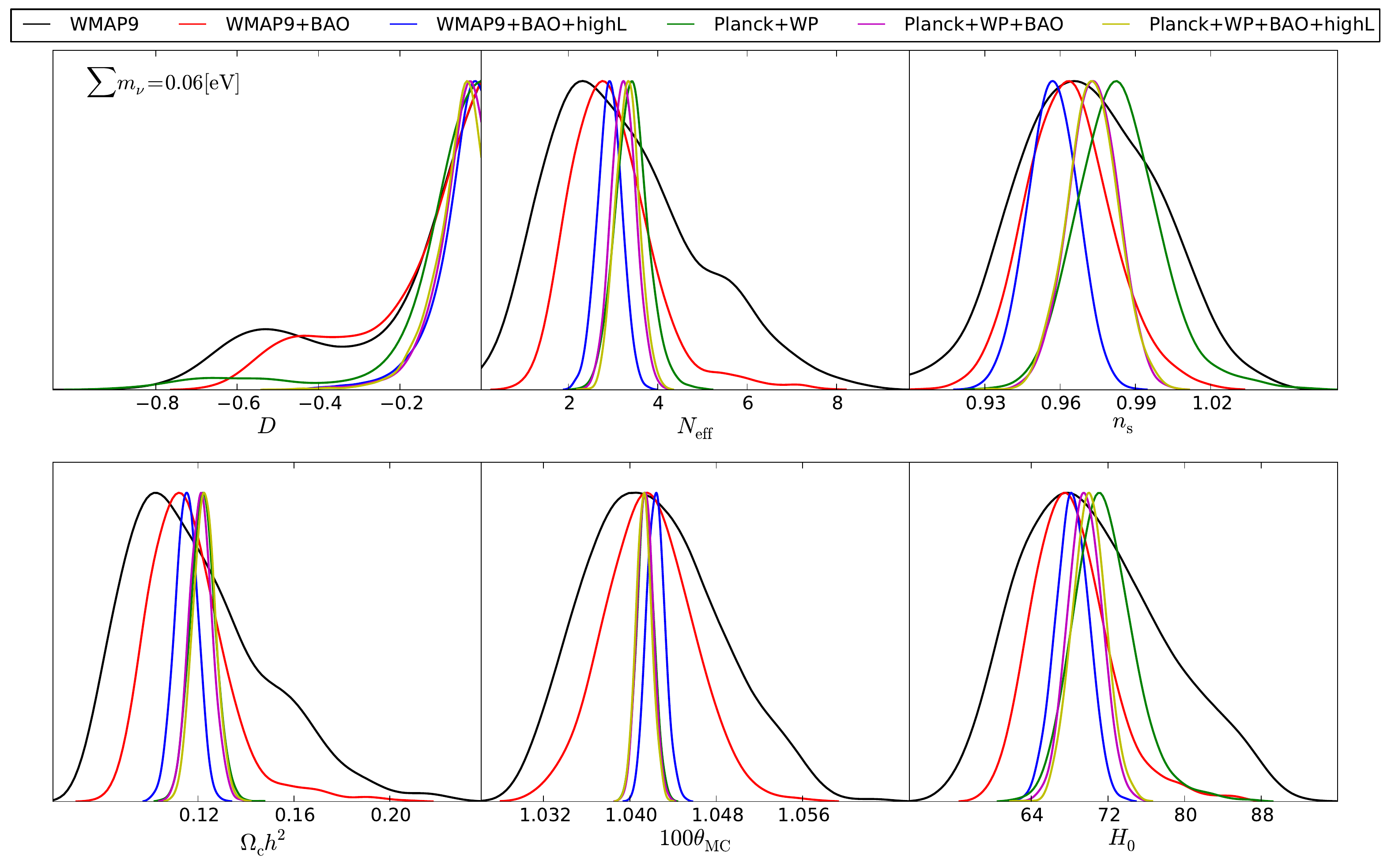}
\caption{One-dimensional marginalized likelihood on the effective number of neutrino species $N_{\rm eff}$ as well as other cosmological parameters $D,\quad n_s,\quad\Omega_ch^2,\quad 100\theta{\rm MC},\quad H_0$. In these $f(R)$ models, we set $\sum m_{\nu}=0.06[{\rm eV}]$.  }\label{neff1D}
\end{figure*}

In this subsection, we consider the constraints on the effective number of neutrino species, $N_{\rm eff}$, assuming the total mass of active neutrinos as $\sum m_{\nu}=0.06{\rm eV}$.
The numerical results are shown in Table \ref{Neff}. In Fig.\ref{neff1D}, we show the one-dimensional marginalized likelihood on the effective number of neutrino species $N_{\rm eff}$ as well as other cosmological parameters $D,\quad n_s,\quad\Omega_ch^2,\quad 100\theta{\rm MC},\quad H_0$. WMAP nine-year data along place rather weak constraints on the effective number of neutrino species
$$N_{\rm eff}=3.28^{+3.33}_{-2.86}(95\%;{\rm WMAP9})$$
at the $95\%$ C.L. However,
the constraints on $N_{\rm eff}$ as well as other cosmological parameters are improved significantly when the BAO data are added. The combination of the WMAP+BAO data set improve the constraint on $N_{\rm eff}$ up to
$$N_{\rm eff}=2.99^{+1.92}_{-1.82}({95\%;\rm WMAP9+BAO})\quad.$$
We find that after adding the high-$l$ data, the constraints can be further improved.
$$N_{\rm eff}=2.92^{+0.53}_{-0.55}(95\%;{\rm WMAP9+BAO+highL})\quad.$$
The error bars have shrunk almost by $50\%$ compared to the case without the high-$l$ data.
The other cosmological parameters are also better constrained after adding the high-$l$ data(see Table \ref{Neff}).
Particularly, $\Omega_ch^2$ is constrained up to $0.1151^{+0.0048}_{-0.0048}$ where the error bars have reduced by almost $75\%$. For the WMAP data set, we can find that the results are compatible with the standard value $N_{\rm eff}=3.046$ within the 1$\sigma$ range.

Compared with the results obtained from the combination of WMAP data, Planck data show robust constraints on $N_{\rm eff}$ as well as the $f(R)$ gravity. Planck data alone in combination with WMAP polarization(WP) data (Planck+WP) give the constraints as
$$N_{\rm eff}=3.43^{+0.76}_{-0.76}(95\%;{\rm Planck + WP})\quad.$$
The best-fit value strongly favors $N_{\rm eff}>3.046$, which indicates the existence of extra species of relativistic neutrinos. The standard value $N_{\rm eff}=3.046$ is only on the edge of the $1\sigma$ range (see Table \ref{Neff}) but is still compatible within the $2\sigma$ range. Adding the BAO data can improve the constraints significantly.
The combination of Planck+WP+BAO data set gives
$$N_{\rm eff}=3.24^{+0.55}_{-0.53}(95\%;{\rm Planck+WP+BAO})\quad.$$
However, we find that further adding the high-$l$ data does not show a significant improvement on the constraint of $N_{\rm eff}$. The combination of
Planck+WP+BAO+highL data sets only give
$$N_{\rm eff}=3.32^{+0.54}_{-0.51}(95\%;{\rm Planck+WP+BAO+highL})\quad,$$
which is almost the same as the result in the $\Lambda$CDM model as reported by Planck team $N_{\rm eff}=3.30^{+0.54}_{-0.51}(95\%{\rm C.L.})$\cite{planck}.
This result is expected because the $f(R)$ models investigated in this work only change the late-time growth history of the Universe and do not change the matter-radiation equality.
If the parameter $\Omega_c$ in the $f(R)$ gravity model is tightly constrained, the constraints on $N_{\rm eff}$, in this case, should be quite close to that in the $\Lambda$CDM model.

\begin{table*}
\caption{Cosmological parameter values for the $f(R)$ models with $\sum m_{\nu}=0.06[{\rm eV}]$. $B_0$ is a derived parameter.  \label{Neff}}
\begin{tabular}{|c||c|c|c|c|c|c|}
\hline
Parameters & WMAP9 & WMAP9+BAO & WMAP9+BAO+highL& Planck+WP & Planck+WP+BAO& Planck+WP+BAO+highL\\
\hline
$\Omega_bh^2$ &$0.02288^{+0.00052}_{-0.00052}$ &$0.02270^{+0.00046}_{-0.00046}$&$0.02252^{+0.00027}_{-0.00027}$&$0.02296^{+0.00048}_{-0.00048}$&$0.02268^{+0.00031}_{-0.00031}$& $0.02269^{+0.00031}_{-0.00031}$ \\
\hline
$\Omega_ch^2$ &$0.1190^{+0.0280}_{-0.0280}$ &$0.1163^{+0.0171}_{-0.0171}$&$0.1151^{+0.0048}_{-0.0048}$&$0.1220^{+0.0052}_{-0.0052}$&$0.1212^{+0.0048}_{-0.0048}$&$0.1226^{+0.0046}_{-0.0046}$\\
\hline
$100\theta_{\rm MC}$ & $1.0422^{+0.0060}_{-0.0060}$
& $1.0418^{+0.0041}_{-0.0041}$&$1.0424^{+0.0008}_{-0.0008}$& $1.0414^{+0.0008}_{-0.0008}$& $1.0414^{+0.0007}_{-0.0007}$&$1.0413^{+0.0007}_{-0.0007}$\\
\hline
$\tau$ &$ 0.0854^{+0.0126}_{-0.0126}$ & $0.0824^{+0.0117}_{-0.0117}$&$0.0793^{+0.0109}_{-0.0109}$& $0.0834^{+0.0120}_{-0.0120}$& $0.0808^{+0.0117}_{-0.0117}$&$0.0810^{+0.0116}_{-0.0116}$ \\
\hline
$n_s$ & $0.9713^{+0.0267}_{-0.0267}$ & $0.9638^{+0.0172}_{-0.0172}$&$0.9570^{+0.0099}_{-0.0099}$ & $0.9840^{+0.0171}_{-0.0171}$& $0.9731^{+0.0099}_{-0.0099}$ &$0.9729^{+0.0101}_{-0.0101}$\\
\hline
$\rm{ln}[10^{10}As]$ & $3.082^{+0.063}_{-0.063}$ &$3.077^{+0.044}_{-0.044}$ & $3.058^{+0.026}_{-0.026}$ &$3.081^{+0.027}_{-0.027}$&$3.074^{+0.025}_{-0.025}$&$3.076^{+0.025}_{-0.025}$\\
\hline
$|D|$ & $ <0.639$(95\%{\rm C.L.}) & $ <0.517$(95\%{\rm C.L.})&$ <0.205$(95\%{\rm C.L.})& $ <0.616$(95\%{\rm C.L.})& $ <0.188$(95\%{\rm C.L.})& $ <0.177$(95\%{\rm C.L.})\\
$(B_0)$ & $ <3.25$(95\%{\rm C.L.}) & $ <2.37$(95\%{\rm C.L.})&$ <0.728$(95\%{\rm C.L.})& $ <3.08$(95\%{\rm C.L.})& $ <0.674$(95\%{\rm C.L.})& $ <0.628$(95\%{\rm C.L.})\\
\hline
$N_{\rm eff}$ & $3.28^{+1.06(+3.33)}_{-2.13(-2.86)}$  & $2.99^{+0.65(+1.92)}_{-1.06(-1.82)}$ &$2.92^{+0.27(+0.53)}_{-0.27(-0.55)}$   &$3.43^{+0.33(+0.76)}_{-0.39(-0.76)}$& $3.24^{+0.27(+0.55)}_{-0.27(-0.53)}$& $3.32^{+0.26(+0.54)}_{-0.27(-0.51)}$\\
\hline
\end{tabular}
\end{table*}

\subsection{Simultaneous constraints on $N_{\rm eff}$ and $\sum m_{\nu}$}
\begin{figure*}
\includegraphics[width=5in,height=4in]{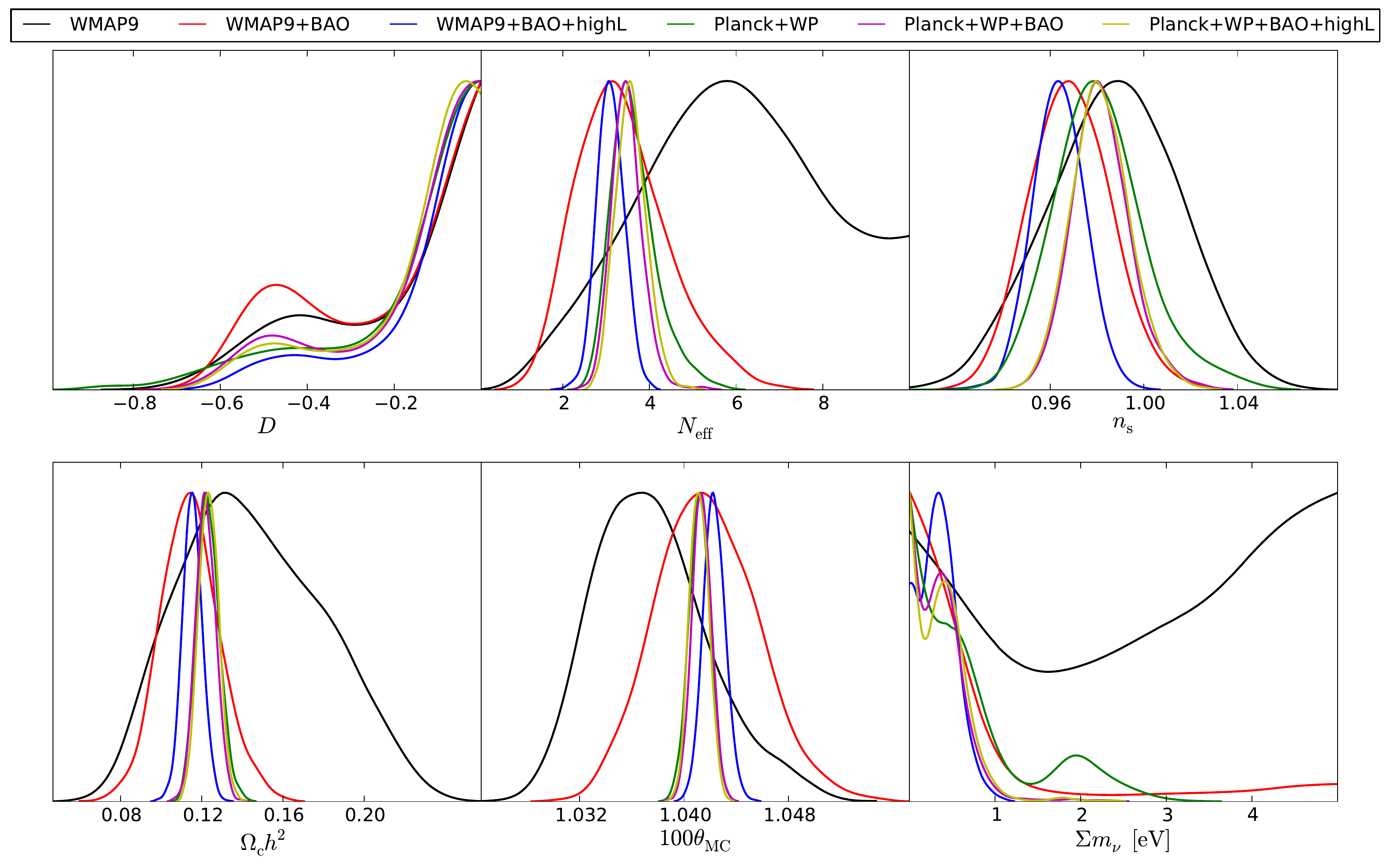}
\caption{One-dimensional marginalized likelihood on $\sum m_{\nu}$, $N_{\rm eff}$ and other cosmological parameters $D,\quad n_s,\quad\Omega_ch^2,\quad 100\theta{\rm MC}$. }\label{Ne_mass1D}
\end{figure*}

In this subsection, we report the joint constraints on the total mass of active neutrinos $\sum m_{\nu}$ and the effective number of species $N_{\rm eff}$. In this work, we assume three active neutrinos share a mass $m_{\nu}=\sum m_{\nu}/3$. The extra species of neutrinos $\delta N_{\rm eff}=N_{\rm eff}-3.046$ are relativistic and massless. When $N_{\rm eff}<3.046$ , the temperature of the three active neutrinos is reduced accordingly, and no additional relativistic species are assumed. Based on these assumptions, we conduct the MCMC analysis and the numerical results are shown in table \ref{Ne_mass}. In Fig.\ref{Ne_mass1D}, we show the one-dimensional marginalized likelihood on $\sum m_{\nu}$, $N_{\rm eff}$ and other cosmological parameters $D,\quad n_s,\quad\Omega_ch^2,\quad 100\theta{\rm MC}$. We first present the results obtained from the data combination associated with WMAP data. WMAP data along yields very poor constraints on both $\sum m_{\nu}$ and $N_{\rm eff}$
\begin{equation}
\left.
\begin{array}{c}
N_{\rm eff} = 5.96^{+4.04}_{-3.42} \\
\sum m_{\nu}< 5{\rm eV}
\end{array}
\right\} \quad(95\%;{\rm WMAP9}).
\end{equation}
The $\sum m_{\nu}$ remains almost unconstrained and the error bars on $N_{\rm eff}$ are quite large. However, these bounds can be significantly tightened by adding BAO data. We find
\begin{equation}
\left.
\begin{array}{c}
N_{\rm eff} = 3.39^{+2.21}_{-1.94} \\
\sum m_{\nu}< 5{\rm eV}
\end{array}
\right\} \quad(95\%;{\rm WMAP9+BAO}).
\end{equation}
However, $\sum m_{\nu}$ still remains almost unconstrained.
After adding the high-$l$ data, we find the constraints are improved significantly.
\begin{equation}
\left.
\begin{array}{c}
N_{\rm eff} = 3.10^{+0.62}_{-0.59} \\
\sum m_{\nu}< 0.712{\rm eV}
\end{array}
\right\} \quad(95\%;{\rm WMAP9+BAO+highL}).
\end{equation}
Similar to previous sections, the Planck data again show robust constraint on both $N_{\rm eff}$ and $\sum m_{\nu}$. We find
\begin{equation}
\left.
\begin{array}{c}
N_{\rm eff} = 3.66^{+1.17}_{-0.99} \\
\sum m_{\nu}< 2.21{\rm eV}
\end{array}
\right\} \quad(95\%;{\rm Planck+WP}).
\end{equation}
However, compared with the results in previous section where $\sum m_{\nu}$ is fixed, the constraint on $N_{\rm eff}$, in this section, is clearly weakened if $\sum m_{\nu}$ can vary. This point is quite different from the case in the $\Lambda$CDM model as reported by the Planck team\cite{planck}, where the joint constraints do not differ very much from the bounds obtained when introducing these parameters separately. This is because $\sum m_{\nu}$ is degenerate with $f(R)$ gravity and looses the constraint on $\Omega_mh^2=\Omega_{\nu}h^2+\Omega_ch^2+\Omega_bh^2$ and so does the matter-radiation equality. The constraint on $N_{\rm eff}$ is, therefore, weakened as well. After adding the BAO data, the constraints are improved up to
\begin{equation}
\left.
\begin{array}{c}
N_{\rm eff} = 3.49^{+0.73}_{-0.71} \\
\sum m_{\nu}< 0.826{\rm eV}
\end{array}
\right\} \quad(95\%;{\rm Planck+WP+BAO}).
\end{equation}
However, we find that adding the high-$l$ data does not show significant improvement on the constraints.
\begin{equation}
\left.
\begin{array}{c}
N_{\rm eff} = 3.58^{+0.72}_{-0.69} \\
\sum m_{\nu}< 0.860{\rm eV}
\end{array}
\right\} \quad(95\%;{\rm Planck+WP+BAO+highL}).
\end{equation}

\begin{table*}
\caption{Cosmological parameter values for the $f(R)$ models with constraining $\sum m_{\nu}$ and
$N_{\rm eff}$ simultaneously. $B_0$ is a derived parameter. \label{Ne_mass}}
\begin{tabular}{|c||c|c|c|c|c|c|}
\hline
Parameters & WMAP9 & WMAP9+BAO & WMAP9+BAO+highL & Planck+WP & Planck+WP+BAO&Planck+WP+BAO+highL\\
\hline
$\Omega_bh^2$ &$0.02324^{+0.00068}_{-0.00068}$ &$0.02296^{+0.00066}_{-0.00066}$&$0.02277^{+0.00032}_{-0.00032}$  &$0.02304^{+0.00059}_{-0.00059}$&$0.02299^{+0.00042}_{-0.00042}$&$0.02299^{+0.00042}_{-0.00042}$ \\
\hline
$\Omega_ch^2$ &$0.1455^{+0.0342}_{-0.0342}$ &$0.1143^{+0.0145}_{-0.0145}$&$0.1154^{+0.0049}_{-0.0049}$&$0.1233^{+0.0056}_{-0.0056}$&$0.1220^{+0.0049}_{-0.0049}$&$0.1234^{+0.0048}_{-0.0048}$\\
\hline
$100\theta_{\rm MC}$ & $1.0375^{+0.0044}_{-0.0044}$
& $1.0417^{+0.0038}_{-0.0038}$&$1.0423^{+0.0008}_{-0.0008}$& $1.0412^{+0.0008}_{-0.0008}$& $1.0413^{+0.0007}_{-0.0007}$&$1.0412^{+0.0007}_{-0.0007}$\\
\hline
$\tau$ &$ 0.0846^{+0.0134}_{-0.0134}$ & $0.0855^{+0.0128}_{-0.0128}$&$0.0821^{+0.0121}_{-0.0121}$& $0.0874^{+0.0138}_{-0.0138}$& $0.0835^{+0.0127}_{-0.0127}$&$0.0860^{+0.0124}_{-0.0124}$ \\
\hline
$n_s$ & $0.9872^{+0.0272}_{-0.0272}$ & $0.9688^{+0.0172}_{-0.0172}$&$0.9638^{+0.0111}_{-0.0111}$ & $0.9815^{+0.0196}_{-0.0196}$& $0.9819^{+0.0126}_{-0.0126}$&$0.9817^{+0.0125}_{-0.0125}$ \\
\hline
$\rm{ln}[10^{10}As]$ & $3.156^{+0.063}_{-0.063}$ &$3.082^{+0.045}_{-0.045}$ &$3.062^{+0.027}_{-0.027}$&$3.093^{+0.033}_{-0.033}$&$3.079^{+0.027}_{-0.027}$&$3.086^{+0.027}_{-0.027}$\\
\hline
$|D|$ & $ <0.565$(95\%{\rm C.L.}) & $ <0.553$(95\%{\rm C.L.})&$<0.490 $(95\%{\rm C.L.}) &$ <0.596$(95\%{\rm C.L.})& $ <0.536$(95\%{\rm C.L.})&$ <0.525$(95\%{\rm C.L.})\\
$(B_0)$ & $ <2.70$(95\%{\rm C.L.}) & $ <2.62$(95\%{\rm C.L.})&$<2.23 $(95\%{\rm C.L.}) &$ <2.92$(95\%{\rm C.L.})& $ <2.50$(95\%{\rm C.L.})&$ <2.43$(95\%{\rm C.L.})\\
\hline
$N_{\rm eff}$ & $5.96^{+2.17(+4.04)}_{-2.30(-3.42)}$  & $3.39^{+0.81(+2.21)}_{-1.27(-1.94)}$ &$3.10^{+0.31(+0.62)}_{-0.33(-0.59)}$ &$3.66^{+0.37(+1.17)}_{-0.63(-0.99)}$& $3.49^{+0.30(+0.73)}_{-0.39(-0.71)}$&$3.58^{+0.33(+0.72)}_{-0.39(-0.69)}$\\
\hline
$\sum m_{\nu}[{\rm eV}]$ & $ <5$(95\%{\rm C.L.}) & $ <5$(95\%{\rm C.L.})&$ <0.712$(95\%{\rm C.L.})& $ <2.21$(95\%{\rm C.L.})& $ <0.826$(95\%{\rm C.L.})& $ <0.860$(95\%{\rm C.L.})\\
\hline
\end{tabular}
\end{table*}

\section{conclusions\label{conclusions}}
In this work, we have analyzed the performance of constraints on neutrino properties from the latest cosmological observations in the framework of $f(R)$ gravity using massive MCMC analysis. We have analyzed the constraints on the total mass of neutrinos $\sum m_\nu$ assuming $N_{\rm eff}=3.046$; we have also analyzed the constraints on the effective number of neutrino species $N_{\rm eff}$ assuming $\sum m_\nu=0.06[{\rm eV}]$;finally,we have analyzed the constraints on $N_{\rm eff}$ and $\sum m_\nu$ simultaneously.

To conclude, we summarize our main results with the tightest error bars in Table\ref{compare} and also compare them with the results obtained by the Planck team\cite{planck} within the context of the $\Lambda$CDM model. We can find that the constraints on $\sum m_{\nu}$ when fixing $N_{\rm eff}=3.046$ in $f(R)$ gravity are  a factor of 2 larger than those of the $\Lambda$CDM model. When fixing $\sum m_{\nu}=0.06{\rm eV}$, the constraint on $N_{\rm eff}$ in $f(R)$ gravity is almost the same as that in the $\Lambda$CDM model. However, when running $\sum m_{\nu}$ and $N_{\rm eff}$ simultaneously, the constraints on $N_{\rm eff}$ and $\sum m_\nu$ in the $f(R)$ model are both significantly weaker than that in the $\Lambda$CDM model due to the degeneracy between the late time growth history in $f(R)$ gravity and $\sum m_{\nu}$.

\begin{table*}
\caption{The comparison of fitting results in the $f(R)$ models and the $\Lambda$CDM model.}\label{compare}
\begin{tabular}{c|c|c}
  \hline
  \hline
 Data &  \multicolumn{2}{c}{ Planck+WP+highL+BAO}    \\
 \hline
  Model & $\Lambda$CDM(95\%{\rm C.L.}) & $f(R)$(95\%{\rm C.L.}) \\
  \hline
 \multicolumn{3}{c}{$N_{\rm eff}=3.046$}    \\
 \hline
 $\sum m_{\nu}$ & $<0.23 {\rm eV}$ & $<0.462{\rm eV}$ \\
  \hline
 \multicolumn{3}{c}{$\sum m_{\nu} =0.06 {\rm eV}$}    \\
 \hline
 $N_{\rm eff}$ & $N_{\rm eff}=3.30^{+0.54}_{-0.51}$ & $N_{\rm eff}=3.32^{+0.54}_{-0.51}$ \\
  \hline

  \multicolumn{3}{c}{{\rm Simultaneous constraints on} $N_{\rm eff}$ and $\sum m_{\nu}$}    \\
  \hline
 $N_{\rm eff}$ & $N_{\rm eff}=3.32^{+0.54}_{-0.52}$ & $N_{\rm eff}=3.58^{+0.72}_{-0.69}$ \\
  \hline
 $\sum m_{\nu}$ & $<0.28{\rm eV}$ & $<0.860{\rm eV}$ \\
 \hline
\end{tabular}
\end{table*}
In summary, constraints on neutrino properties from cosmological observations are highly model dependent. Tighter constraints on the neutrino properties can only be achieved when the modified gravity models are also well constrained.

Stringent constraints on the $f(R)$ model can be obtained on nonlinear scales using the data from cluster abundance\cite{cluster}. However, the chameleon mechanism\cite{Mota,Khoury} plays an important role on nonlinear scales. At early times, since the background curvature is very high, the nonliner perturbation for the $f(R)$ models which can go back to the $\Lambda$CDM model at high curvature regime $\lim_{R\rightarrow+\infty}F(R)=1$ generally follows the "high-curvature solution" \cite{HuI}, where the effective Newtonian constant in overdensity regions is extremely close to that of the standard gravity $G_{\rm eff}\sim G$\cite{Hesim} and the chameleon mechanism works very efficiently in this period. If the "high-curvature solution" in high density regions could persist until present day, the thin-shell structure can be formed naturally in high density regions for the galaxies in the Universe. If the galaxies are sufficiently self-screened, the stars inside a galaxy can naturally be self-screened as well. The model thus can evade the stringent local tests of gravity. However, in high density regions, the "high-curvature solutions" are not always achieved for $f(R)$ models at late times in the Universe.
For the family of $f(R)$ models studied in this work, neglecting the effects of massive neutrinos, we do not find any "high-curvature solutions" or "thin-shell" structures in the dense region for the models with $|f_{R0}=1-F|>10^{-4}$ and there is a factor of $1/3$ enhancement in the strength of Newtonian gravity\cite{Hesim}. This means that these models could be ruled out by local tests of gravity and, conservatively speaking, the viable $f(R)$ models should be with $|f_{R0}=1-F|<=10^{-4}(B_0<5.5\times 10^{-4})$~\cite{simulation}. From the tightest  astrophysical constraints $B_0<2.5\times10^{-6}$\cite{Jain} which is in the bound of $B_0<5.5\times 10^{-4}$, we can learn that, for viable $f(R)$ models, at least, the chameleon screening mechanism should work very efficiently. However, this estimation is only based on our simulations in the case without taking account of massive neutrinos. There are no N-body simulations available at the moment, to our best knowledge, that have been calibrated with neutrinos in any forms of $f(R)$ models. To calibrate neutrinos in $f(R)$ simulations is an urgent object of our future work.

\emph{Acknowledgment: J.H.He acknowledges the Financial
support of MIUR through PRIN 2008 and ASI through
contract Euclid-NIS I/039/10/0. We thank B. R. Granett for carefully reading the manuscript.}

\end{document}